\documentclass[showkeys,aps,prl,amsmath,amssymb,reprint,floatfix,superscriptaddress]{revtex4-2}

\usepackage[caption=false]{subfig}
\usepackage[normalem]{ulem}
\usepackage{lipsum}
\usepackage[linkcolor=red, citecolor=blue, colorlinks=true, urlcolor=blue]{hyperref}
\usepackage{graphicx}
\usepackage{dcolumn}
\usepackage{bm}
\usepackage{xcolor}
\usepackage{xspace}

\usepackage{tikz}
\usetikzlibrary{patterns}
\usetikzlibrary{calc}
\usetikzlibrary{decorations.pathreplacing,decorations.markings}
\usetikzlibrary{arrows}

\tikzset{
  on each segment/.style={
    decorate,
    decoration={
      show path construction,
      moveto code={},
      lineto code={
        \path [#1]
        (\tikzinputsegmentfirst) -- (\tikzinputsegmentlast);
      },
      curveto code={
        \path [#1] (\tikzinputsegmentfirst)
        .. controls
        (\tikzinputsegmentsupporta) and (\tikzinputsegmentsupportb)
        ..
        (\tikzinputsegmentlast);
      },
      closepath code={
        \path [#1]
        (\tikzinputsegmentfirst) -- (\tikzinputsegmentlast);
      },
    },
  },
  mid arrow/.style={postaction={decorate,decoration={
        markings,
        mark=at position .65 with {\arrow[#1]{stealth}}
      }}},
}

\newcommand{\sect}[1]{\vspace{0.2cm}    \noindent {\it{#1} --\xspace}}
\newcommand{\fref}[1]{Fig.~\ref{#1}}

\newcommand{\ie}{{\it i.e.}\xspace}
\newcommand{\eg}{{\it e.g.}\xspace}
\newcommand{\p}{\partial}
\newcommand{\vx}{{\bf x}}
\newcommand{\vq}{{\bf q}}
\newcommand{\vp}{{\bf p}}
\newcommand{\vv}{{\bf v}}
\newcommand{\vepsilon}{{\bf e}}
\newcommand{\vnabla}{{\bf \nabla}}

\begin{document}

\title[]{The unpredicted scaling of the one-dimensional Kardar-Parisi-Zhang equation}

\author{C\^ome Fontaine}
\affiliation{Univ. Grenoble Alpes, CNRS, LPMMC, 38000 Grenoble, France}
\author{Francesco Vercesi}
\affiliation{Univ. Grenoble Alpes, CNRS, LPMMC, 38000 Grenoble,  France}
\author{Marc Brachet}
\affiliation{Universit\'e PSL, CNRS, Sorbonne Universit\'e, Universit\'e de Paris, LPENS, 75005, Paris}
\author{L\'eonie Canet}
\affiliation{Univ. Grenoble Alpes, CNRS, LPMMC, 38000 Grenoble, France}
\affiliation{Institut Universitaire de France, 75000 Paris, France}

\begin{abstract}
The celebrated Kardar-Parisi-Zhang (KPZ) equation describes the kinetic roughening of stochastically growing interfaces. In one dimension, the KPZ equation is exactly solvable and its  statistical properties  are known to an exquisite degree. Yet recent numerical simulations in the tensionless (or inviscid) limit of the KPZ equation [Phil. Trans. Roy. Soc. A {\bf 380} 20210090 (2022), Phys. Rev. E {\bf 106} 024802 (2022)] unveiled a new scaling, with a critical dynamical exponent $z=1$  different from the KPZ one $z=3/2$. In this Letter, we show that this scaling is controlled by a fixed point which had been missed so far and which corresponds to an infinite non-linear coupling. Using the functional renormalization group (FRG), we demonstrate the existence of this fixed point and show that it yields $z=1$. We  calculate the correlation function and  associated scaling function at this fixed point, providing both a numerical solution of the FRG equations within a reliable approximation, and an exact asymptotic form obtained in the limit of large wavenumbers. Both scaling functions accurately match the one from the numerical simulations.
\end{abstract}

\maketitle

The Kardar-Parisi-Zhang (KPZ) equation is remarkable for the large variety of systems in which it arises. 
 Originally derived to model the kinetic roughening of stochastically growing interfaces \cite{Kardar86}, the KPZ equation  has turned out to describe the universal properties of systems as different as various growing interfaces \cite{Maunuksela97,Huergo2012,Wakita97,Najem2020},
   equilibrium disordered systems  \cite{kardar87}, or turbulence in infinitely compressible fluids \cite{Burgers48}. Perhaps even more striking is its recent observation in purely quantum systems, such as exciton-polariton condensates \cite{Fontaine2022Nat} or Heisenberg quantum spin chains \cite{Bloch2022,Tennant2022}. The ubiquity of the KPZ equation promotes it to a fundamental model for non-equilibrium critical phenomena and phase transitions  \cite{Halpin-Healy95,Krug97,Halpin-Healy15,Takeuchi18}.

After more than two decades of intense efforts both in the mathematics and statistical physics communities, the one-dimensional (1D) KPZ equation has been solved exactly, and its statistical properties are now extensively charted \cite{Corwin12}. In 1D, the critical exponents of the KPZ equation, roughness exponent $\chi$ and dynamical exponent $z$, are fixed by the symmetries to the exact values $\chi=1/2$ and $z=3/2$.  The two-point correlation function has  been calculated exactly \cite{Praehofer04}.
The probability distribution of the KPZ height fluctuations is  known, and reveals a sensitivity to the global geometry of the interface,  while  unveiling a deep connection with random matrix theory \cite{Corwin12}. 
Many other properties are also known, such as the short-time behavior or the large deviation theory, to cite a few. However, the 1D KPZ equation still reserves its surprises. 

In a recent paper \cite{Brachet2022}, the authors performed numerical simulations of the 1D Burgers equation \cite{Bec2007}, which exactly maps to the KPZ equation, and studied the limit of vanishing viscosity (inviscid limit). They unveiled a crossover to a new scaling regime, characterized by a dynamical exponent $z=1$, different from the KPZ value $z=3/2$. The same result was reported in \cite{Rodriguez2022} in the equivalent tensionless limit of the KPZ equation, and also in \cite{Fujimoto2020} in a strongly interacting 1D quantum bosonic system.  This scaling is absent in the current understanding of the 1D KPZ equation. In this Letter, we fill this gap, and provide the theoretical explanation of this missing scaling, using the functional renormalization group (FRG). In the renormalization group framework, the KPZ scaling is controlled by a fixed point, termed the KPZ fixed point. Another fixed point exists, the Edwards-Wilkinson (EW) fixed point, which corresponds to the KPZ equation with  vanishing non-linearity \cite{Edwards82}. We show that the tensionless or inviscid limit of the KPZ equation is controlled by a third unexplored fixed point, which features  the $z=1$ critical dynamical exponent. We calculate the scaling function at this fixed point, and show that it  very accurately coincides with the scaling function computed in the numerics.

Let us first justify on simple grounds the existence of this third fixed point.   The KPZ equation gives the dynamics of a real-valued height field $h(t,\vx)$ with $\vx\in \mathbb{R}^d$: 
\begin{equation}
\p_t h = \nu \nabla^2 h + \dfrac \lambda 2 \big(\vnabla h\big)^2 + \sqrt{D}\eta
 \label{eq:kpz}
\end{equation}
where $\nu$, $\lambda$ and $D$ are three real parameters and $\eta$ is a Gaussian noise of zero mean and correlations $\langle\eta(t,\vx)\eta(t',\vx')\rangle=2 \delta(t-t')\delta^d(\vx-\vx')$. In fact, by rescaling the time and the field, one can show that this equation only depends on one parameter $g\equiv \lambda^2 D/\nu^3$ (or equivalently on the Reynolds number in the context of the Burgers equation). 
 Note that we assume, as in \cite{Brachet2022}, the existence of an UV cutoff scale,  such that the solutions of \eqref{eq:kpz} remain well-defined in the inviscid limit  and thermalize to the equilibrium distribution \cite{Majda2000,Brachet2022}.

Within the RG framework, following Wilson's original idea, one progressively averages out fluctuations, shell by shell in wavenumbers, starting from the high (ultraviolet UV) wavenumber modes \cite{Wilson74}. One thus obtains the effective theory for the low (infrared IR) wavenumber modes, \ie at large distances. When the system is scale invariant, this corresponds to a fixed point of the RG flow. Thus, the KPZ rough interface is described by an IR fixed point, the KPZ one, which is fully attractive in 1D and is characterized by $z=3/2$ and $\chi=1/2$. 

\begin{figure}[h]
\includegraphics[width=8cm]{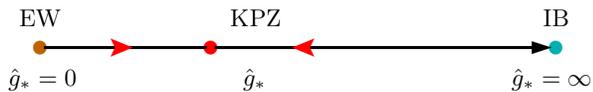}
\caption{The three fixed points of the KPZ equation, the KPZ one, which is IR stable, and the EW and IB ones, which are both IR unstable, UV stable. Red arrows indicate the RG flow.}
\label{fig:schem-g}
\end{figure}

At zero non-linearity $\lambda=g=0$, the  equation \eqref{eq:kpz} becomes the EW equation, and  there exists the corresponding fixed point describing the linear system, characterized  in 1D by $z=2$ and $\chi=1/2$. This fixed point is repulsive, \ie IR unstable. This is schematically depicted in \fref{fig:schem-g}.
From a topological view-point, it is clear that there should also exist another fixed point, in the limit $g\to \infty$, which is IR unstable (and UV stable). Moreover, let us emphasize that the inviscid limit is equivalent to the  limit $g\to \infty$. It is therefore plausible that this third fixed point governs the KPZ equation in this limit, and we call it the inviscid Burgers (IB) fixed point. We show in this paper that it is indeed the case, and that this fixed point yields $z=1$. Since it is genuinely non-perturbative, a  method such as the FRG is required to study it.

\sect{Functional Renormalization Group for KPZ}

The FRG is a modern and powerful implementation of the RG, which allows for both functional and non-perturbative calculations \cite{Wetterich93}, and is widely used in many domains \cite{Berges2002,Dupuis2021}. For the KPZ equation, the FRG yields the strong-coupling fixed point describing the KPZ rough phase in any dimension \cite{Canet2010}, whereas perturbation theory, even resummed to all orders, fails to access it in $d\geq 2$ \cite{Wiese98}. In 1D, the scaling function associated with the two-point correlation function calculated from FRG compares at a high precision level with the exact result  \cite{Canet2011kpz}. Moreover, it can be extended to arbitrary dimensions where it allowed for the calculation of the scaling function and other properties  in $d>1$ \cite{Kloss2012,Kloss2014a,Squizzato2019}.  We thus employ  this method to investigate the IB fixed point.

The starting point of the FRG is the KPZ field theory, which can be obtained from a standard  procedure  introducing a response field $\tilde{h}$ \cite{Martin73,Janssen76,Dominicis76,Langouche79}, and reads
\begin{align}
{\cal Z}[{\cal J}] &= \int {\cal D}h {\cal D}\tilde{h} \; e^{-{\cal S}_{\rm KPZ}[\varphi]+\int_{t,\vx} {\cal J}\cdot{\cal \varphi}} \,\label{eq:actionKPZ}\\
{\cal S}_{\rm KPZ}[\varphi] &= \int_{t,\vx}\Big\{\tilde{h}\Big[\p_t h -\nu \nabla^2 h - \dfrac \lambda 2 \big(\vnabla h\big)^2   \Big] - D\tilde{h}^2\Big\}\, ,
\nonumber
\end{align}
where $\varphi =(h,\tilde{h})$, ${\cal J} =(J,\tilde{J})$ are the sources, and $\int_{t,\vx}\equiv \int dtd^d\vx$
\footnote{Note that we implicitly used It$\bar{\rm o}$'s discretization such that the functional  determinant is unity (see App.~\ref{app:FRG}).}.
The FRG formalism consists in {\it progressively} integrating  the fluctuations in ${\cal Z}$, by suppressing the contribution of low wavenumber modes $q=|\vq\,|\lesssim \kappa$, where $\kappa$ is the RG scale, in the functional integral. This is achieved by adding to ${\cal S}_{\rm KPZ}$ a quadratic  term of the form
$\Delta{\cal S}_\kappa[\varphi] = \frac 1 2 \int \, \varphi_i {\cal R}_{\kappa,ij}\varphi_j\,$,  where ${\cal R}_{\kappa}$ is a $2\times 2$ matrix, whose elements ${\cal R}_{\kappa,ij}$ are called cutoff functions or regulators. They are required to be large  ${\cal R}_{\kappa,ij}(\vq\,)\sim \kappa^2$ at low wavenumbers $q\lesssim\kappa$ such that these modes are damped in the functional integral, and to vanish ${\cal R}_{\kappa,ij}(\vq\,)=0$ at high wavenumbers $q\gtrsim\kappa$ such that these modes are unaffected. Its precise form is unimportant (we refer to App.~\ref{app:FRG} for technical details).
In the presence of $\Delta{\cal S}_\kappa$, ${\cal Z}$ becomes $\kappa$-dependent, and one defines the effective average action $\Gamma_\kappa$, as the Legendre transform of ${\cal W}_\kappa\ = \ln{\cal Z}_\kappa$, {\it i.e.}
$\Gamma_\kappa = -{\cal W}_\kappa + \int_{t,\vx} {\cal J}\cdot \psi -\Delta{\cal S}[\psi]$,
where $\Psi =(\psi,\tilde{\psi})= \big\langle\varphi\big\rangle$. The $\Delta{\cal S}[\psi]$ term, with  
the requirement that the cutoff functions diverge at the microscopic scale $\kappa=\Lambda$ and vanish at $\kappa=0$,  ensures that $\Gamma_\kappa$
 identifies with the microscopic KPZ action \eqref{eq:actionKPZ} at $\kappa=\Lambda$, and becomes the full $\Gamma$, which encompasses all the statistical properties of the system, in the limit $\kappa\to 0$. The evolution of $\Gamma_\kappa$ with the RG scale in between these two scales is given by the Wetterich exact RG equation \cite{Wetterich93}
 \begin{equation}
 \partial_\kappa \Gamma_\kappa = \dfrac 1 2 {\rm Tr}\int\, \p_\kappa{\cal R}_\kappa \cdot  G_\kappa\, ,\quad G_\kappa\equiv \Big[\Gamma_\kappa^{(2)} + {\cal R}_\kappa\Big]^{-1}\, ,
 \label{eq:dsGamkmain}
 \end{equation}
where $\Gamma_\kappa^{(2)}$ is the Hessian of $\Gamma_\kappa$.
The power of the FRG formalism is that this equation can be solved using non-perturbative and functional approximation schemes \cite{Dupuis2021}.

\sect{Flow diagram of the 1D KPZ equation}

Let us  confirm the existence of the IB fixed point. For this, the simplest approximation, which consists in  considering the flow of the original parameters $\nu$, $\lambda$ and $D$ only, suffices. The corresponding ansatz for $\Gamma_\kappa$ reads
\begin{equation}
\Gamma_\kappa = \int_{t,\vx} \Big\{ \tilde{\psi}\Big[\p_t \psi -\nu_\kappa \nabla^2 \psi - \dfrac{\lambda_\kappa}{2} \big(\vnabla \psi\big)^2   \Big] - D_\kappa\tilde{\psi}^2\Big\}\, .
\label{eq:anzLPAmain}
\end{equation}	
From this ansatz, one deduces, measuring in units of $\kappa$, that the frequency scales as $\omega\sim\kappa^2 \nu_\kappa$, and that the fields have scaling dimensions $[\tilde\psi] =\big(\kappa^{d+2} D_\kappa^{-1}\nu_\kappa\big)^{1/2}$ and $[\psi] =\big(\kappa^{d-2} D_\kappa\nu_\kappa^{-1}\big)^{1/2}$. One can show that the coupling $\lambda$ is not renormalized, \ie $\lambda_\kappa = \lambda$ at all scales,  due to the statistical tilt symmetry of the KPZ equation, or equivalently the Galilean invariance of the Burgers equation \cite{Frey94,Canet2011kpz}. Defining the anomalous dimensions $\eta_\kappa^\nu=-\p_s \ln \nu_\kappa$ and  $\eta_\kappa^D=-\p_s \ln D_\kappa$, with $s=\ln(\kappa/\Lambda)$ the RG `time', one deduces that the critical exponents are obtained as $z=2-\eta_*^\nu$ and $\chi = (2-d -\eta_*^\nu+\eta_*^D)/2$, where  $_*$ denotes fixed-point values \cite{Canet2010}.
One then defines the dimensionless effective coupling
$\hat{g}_\kappa  = \kappa^{d-2}\lambda^2 D_\kappa/\nu_\kappa^3$. Its flow equation is given by 
 $\p_s \hat{g}_\kappa = \hat{g}_\kappa \big(d-2-\eta^D_\kappa + 3 \eta^\nu_\kappa \big)$.
The expressions of $\eta_\kappa^\nu$ and $\eta_\kappa^D$ are obtained from projecting the exact flow equation \eqref{eq:dsGamkmain} onto the ansatz \eqref{eq:anzLPAmain}. The calculation is detailed in App.~\ref{app:NLO}.
 At a finite fixed point $0<\hat{g}_*<\infty$, one thus finds the exact identity $z+\chi=2$, whereas the exponent values are not constrained if $\hat{g}_*$ vanishes or diverges. Moreover, in 1D, the accidental time-reversal symmetry further imposes 
 that $D_\kappa=\nu_\kappa$ \cite{Frey94,Canet2011kpz}, and thus $\eta^D_\kappa =\eta^\nu_\kappa \equiv\eta_\kappa$, which leads to $\chi=\eta_*=1/2$. We require this symmetry to be preserved for all values of $\nu$, such that the inviscid limit corresponds to a joined limit $\nu\to0$, $D\to0$ with $\nu/D$ fixed, as in \cite{Brachet2022}. This yields that the stationary solution is a Brownian in space and $\chi=1/2$ for all $\nu$. The flow  equation for $\hat{g}_\kappa$ also possesses the two fixed-point solutions $\hat{g}_*=0$ and $\hat{g}_*=\infty$. In order to render this more explicit, let us change variable to $\hat{w}_\kappa = \hat{g}_\kappa/(1+\hat{g}_\kappa)$. The flow equation for $\hat{w}_\kappa$ reads
$ \p_s \hat{w}_\kappa = \hat{w}_\kappa\,(1-\hat{w}_\kappa) \, \big(2  \eta_\kappa -1 \big)\,$. 
The explicit equation for $\eta_\kappa$ can be found in App.~\ref{app:FRG}, which shows that it is vanishes for $\hat{w}_\kappa=0$.
It is manifest that this equation possesses the three following fixed-point solutions:
i) EW with  $\hat{w}_*=0$, $\eta_*=0$ and thus $z_{\rm \tiny EW}=2$, ii) KPZ with  $0<\hat{w}_*<1$,  $\eta_*=1/2$ and thus $z_{\rm\tiny KPZ}=3/2$, iii) IB with $\hat{w}_*=1$. However, $\eta_*$ is not fixed in this case by the fixed point condition $\p_s \hat{w}_\kappa=0$ and has to be calculated from the flow.
 While it provides the confirmation of the scenario schematically depicted on \fref{fig:schem-g}, this  simple approximation is not sufficient to reliably conclude on the value of $z_{\rm \tiny IB}$ (see App.~\ref{app:FRG}). We now show how to determine this value.

\sect{FRG flow equations within the NLO approximation.}

\begin{figure}[t]
\includegraphics[width=7.5cm]{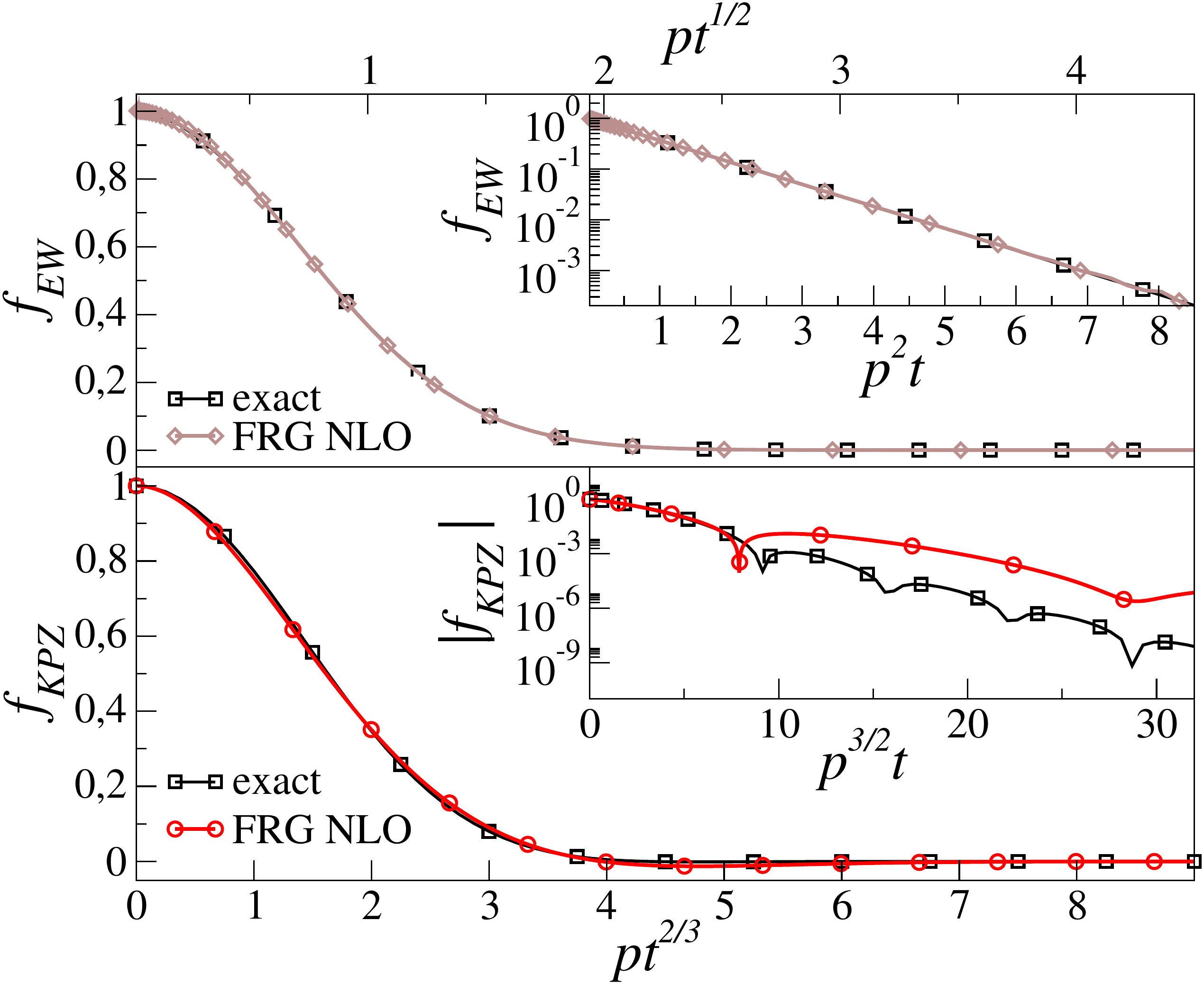}
\includegraphics[width=7.5cm]{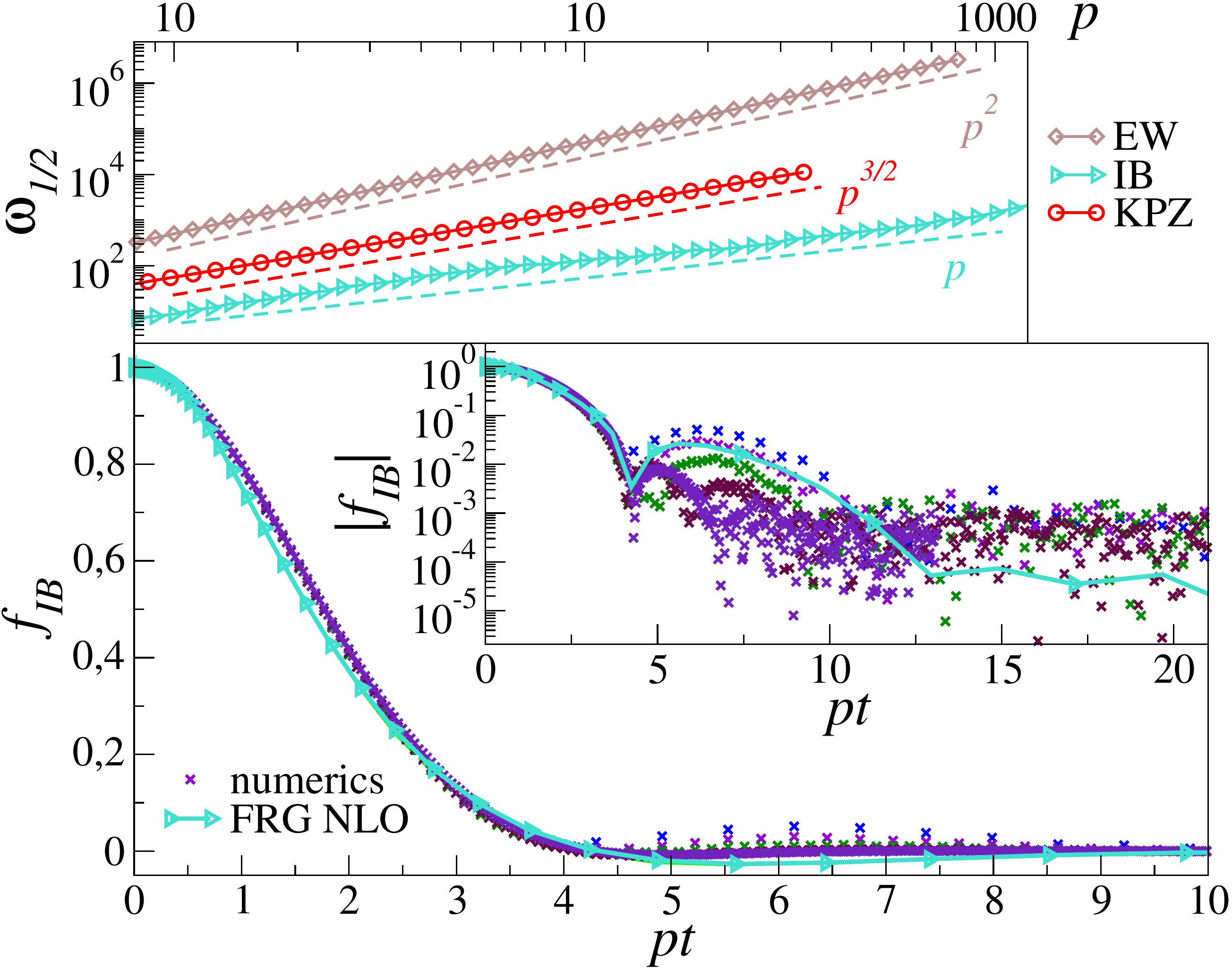}
\caption{Results from the numerical integration of the FRG flow equations within the NLO approximation, obtained from the UV flow at either small (EW) or large (IB) initial coupling  $\hat g_\Lambda$, and from the IR fixed point (KPZ). Half-frequency $\omega_{1/2}$ (shifted vertically for visibility) as a function of $p$, showing the 3 dynamical scaling exponents $z=3/2$ for KPZ, $z=2$ for EW and $z=1$ for IB.  Associated scaling functions $f_{KPZ}$,  $f_{EW}$ and $f_{IB}$, compared respectively with the exact results from \cite{Praehofer04}, with the analytical solution recalled in  App.~\ref{app:NLO}, and with numerical simulations from \cite{Brachet2022}, respectively.}
\label{fig:scaling-function}
\end{figure}

In order to have a quantitative description of the three fixed points, we resort to a refined approximation, introduced in  \cite{Kloss2012} and called next-to-leading-order (NLO) approximation. The NLO ansatz consists in replacing in \eqref{eq:anzLPAmain} the effective parameters $\nu_\kappa$ and $D_\kappa$ by full effective functions $f^\nu_\kappa(\omega,p)$ and $f^D_\kappa(\omega,p)$ respectively. In 1D, the time-reversal symmetry imposes $f^\nu_\kappa=f^D_\kappa\equiv f_\kappa(\omega,p)$, and there is only one anomalous dimension $\eta_\kappa = -\p_s\ln D_\kappa$. 
 The corresponding NLO  equations are derived in App.~\ref{app:NLO}. We numerically solve the flow equation for the dimensionless function $\hat{f}_\kappa(\hat \omega,\hat p) = f_\kappa(\omega/(\kappa^2 D_\kappa),p/\kappa)/D_\kappa$, together with the equation for $\hat g_\kappa$ and $\eta_\kappa$, on a discretized grid $(\hat \omega, \hat p)$ starting from the initial condition  $\hat{f}_\Lambda(\hat \omega,\hat p)=1$ and $\hat g_\Lambda$ at the microscopic scale $\kappa=\Lambda$. The numerical integration is detailed  in App.~\ref{app:FRG}.  For any initial value $\hat g_\Lambda$, the flow reaches in the IR the KPZ fixed point. One can compute from it the correlation function $\bar{C}(\varpi,p)={\cal F}\big[\big\langle\big(h(t',\vx')-h(t,\vx)\big)^2 \big\rangle\big]$ (with ${\cal F}$ the Fourier transform) as 
$\bar{C}(\hat \varpi,\hat p) = {2 {\hat{f}_*}(\hat\varpi, \hat p)}/\big({\hat\varpi^2 + \hat p^4 {\hat{f}_*}^2(\hat\varpi, \hat p)}\big)$.
 The dynamical exponent can be probed through the half-frequency, defined as 
 $\hat{C}(\hat{\varpi}_{1/2}(p),\hat{p}) =  \hat{C}(0,\hat{p})/2$, which shows the expected KPZ scaling $\varpi_{1/2}\sim \hat p^{3/2}$. Fourier transforming back in time the correlation function, one obtains that the data
 for $C(\hat t,\hat p)/C(0,\hat p)$ all collapse onto a single curve when plotted as a function of $p t^{2/3}$, which defines the universal KPZ scaling function. 
 We show in Fig. \ref{fig:scaling-function} that the NLO scaling function compares  accurately with the exact result from \cite{Praehofer04}. It reproduces in particular the negative dip followed by a stretched exponential tail with superimposed oscillations.

Although the flow always reaches in the IR the KPZ fixed point, irrespectively of the initial value of $\hat g_\Lambda$, the beginning of the flow, referred to  as the UV flow, is sensitive to it.  For  small initial values 
$\hat g_\Lambda \ll \hat{g}_*$,  the UV flow is dominated by the  EW  fixed point, while for large $\hat g_\Lambda \gg \hat{g}_*$, it is controlled by the IB one. We  compute the corresponding correlation functions, half-frequency, and scaling functions by focusing on the UV flow starting from either  $\hat g_\Lambda = 10^{-4}$ or $\hat g_\Lambda = 10^{4}$. The results are displayed in  Fig. \ref{fig:scaling-function}. The half-frequency clearly shows two other scaling regimes besides the KPZ one, which are $z=2$ for EW, and $z=1$ for IB. The EW scaling function exactly matches the expected result (recalled in App.~\ref{app:NLO}) up to numerical precision. The IB scaling function is in close agreement with the data from the simulations of \cite{Brachet2022}, at least for $pt\lesssim 4$, featuring in particular the observed negative dip. This confirms that the IB fixed point indeed yields a critical exponent $z=1$. We have thus unveiled the theoretical origin of the missing scaling.

\sect{Exact asymptotic form of the IB scaling function.}

The previous results were derived within the NLO approximation of the FRG. We now show that we can in fact prove the $z=1$ scaling in the UV, and obtain an exact asymptotic form of the scaling function, by considering the limit of large wavenumber $p$.
The proof  is in close analogy with the derivation presented in \cite{Canet2016,Tarpin2018,Canet2022} for the Navier-Stokes (NS) equation. In this case, it was shown that the flow equation for any $n$-point correlation  function $C^{(n)}$ can be closed exactly in the limit of large wavenumbers $p_i=|\vp_i|$. The closure relies on two fundamental ingredients. The first one is the presence of  $\p_\kappa{\cal R}_\kappa$ in the exact FRG flow equations, which ensures that they can be safely expanded in the limit of large $p_i$ (see App.~\ref{app:largep} for details). The second one is the existence of extended symmetries, which exactly fix the expression of the expanded vertices entering the flow equation at large $p_i$. Moreover, it turns out that the resulting closed flow equation for any $C^{(n)}$ can be  solved at the fixed point. This solution gives the exact time dependence of  $C^{(n)}(\{t_i,\vp_i\}_{i=1,n})$
 in the limit of large $p_i$ \cite{Tarpin2018}. 
  These results were precisely confirmed for the two- and three-point functions by direct numerical simulations \cite{Gorbunova2021,Gorbunova2021scalar}. Moreover,  these simulations showed that the regime of validity of the large $p$ expansion starts at wavenumbers larger, but not too far from the inverse integral scale, which means that  it encompasses wavenumbers within the universal inertial range down to the dissipative range. 
 
 To simply exploit the analogy with the NS case, let us consider the action for the Burgers equation in 1D
 \begin{equation*}
 {\cal S}_{\rm Burgers}  \!\! = \!\! \int_{t, x}  \Big\{ \bar{v} {\Big[\p_t v + v \p_x v -\nu \,\p_x^2 v \Big]}   - {D \big(\p_x\bar{v}\big)^2}\, \Big\}\,,
\end{equation*}
where the form of the noise follows from the mapping with the KPZ equation \footnote{Note that the Burgers equation was also studied using FRG in \cite{Mathey2014}.}\cite{Mathey2014}.
 This action shares with the NS one an extended symmetry which is the time-dependent Galilean symmetry: $(t,\vx,\vv) \to (t,\vx + \vepsilon(t),\vv -\dot{\vepsilon}\,(t))$, where $\vepsilon(t)$ is an arbitrary infinitesimal time-dependent vector.
Indeed, this transformation does not leave the Burgers or NS actions strictly invariant, but their variations are {\it linear} in the fields. In such a case, one can derive exact relations, called Ward identities, amongst the vertices $\Gamma^{(n)}$.  The Ward identities for the Galilean extended symmetry entail that each $n$-point vertex with a zero-wavevector associated with a velocity field is exactly given in terms of $(n-1)$-point vertices \cite{Canet2015,Tarpin2018} and App.~\ref{app:largep}.
 It turns out that in 1D, the Burgers action also admits a time-dependent shift symmetry, which is simply $\bar v \to \bar v +\bar{e}(t)$. This only holds in 1D because the advection term can be written in this dimension only as a total derivative. This extended symmetry also yields a set of exact Ward identities, which entail that  each $n$-point vertex with a zero-wavevector associated with a response velocity field exactly vanishes. All these identities are explicitly derived in App.~\ref{app:largep}.
 
\begin{figure}[t]
\includegraphics[width=7.5cm]{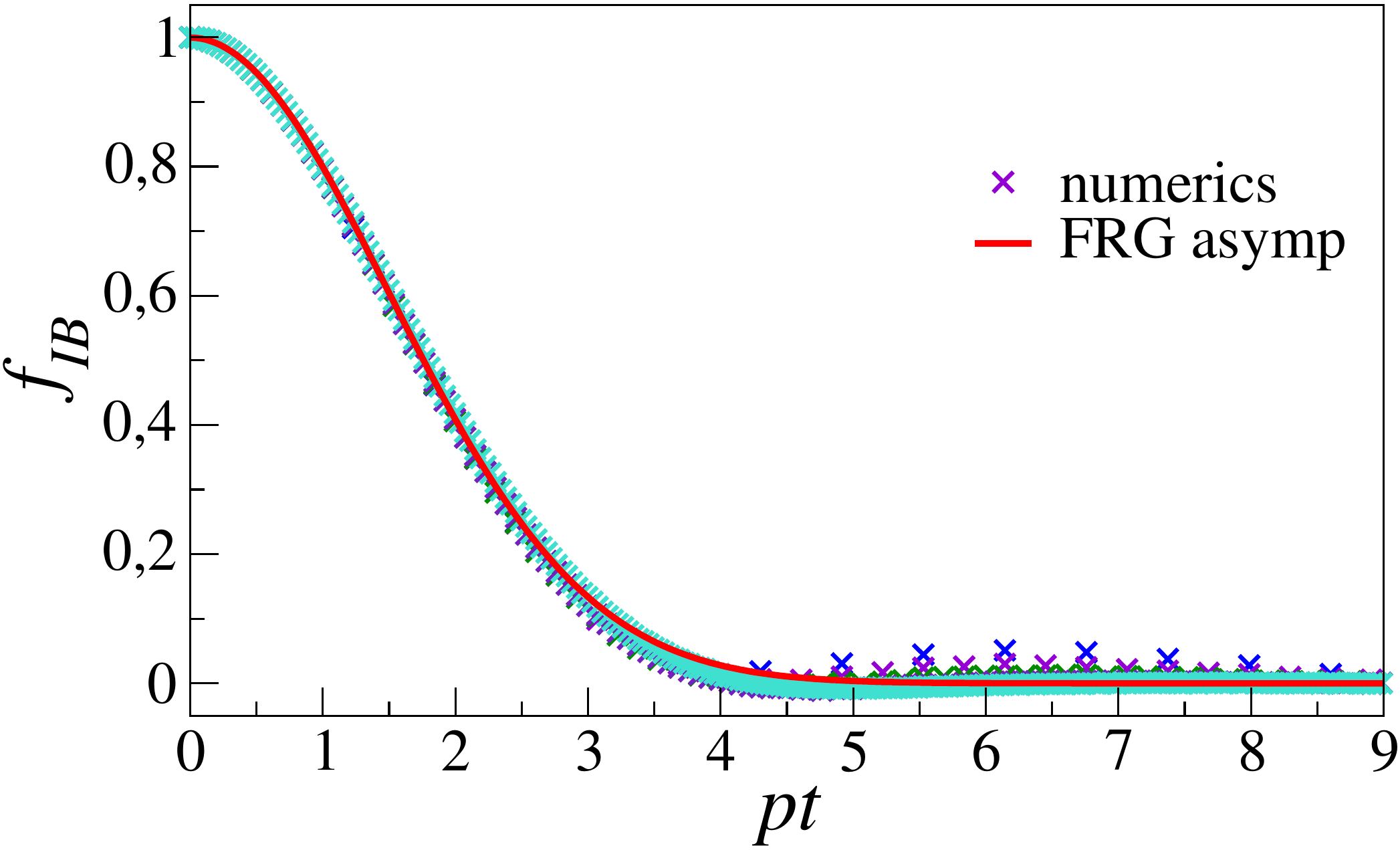}
\caption{Asymptotic form of the  IB scaling function obtained from the solution of the exact FRG flow equation at large wavenumber  and small time, compared with  the  numerics from \cite{Brachet2022}.}
\label{fig:scaling-function-IB}
\end{figure}

 After a calculation, reported in App.~\ref{app:largep}, which is lengthy but very similar to \cite{Canet2016}, one obtains that the flow equation for the two-point function $C_\kappa(t,p)$ is exactly closed in the limit of large $p$. Moreover, it can be solved  at the fixed point, leading to the explicit form
 \begin{equation}
C(t,p) = C(0,p)\times \left\{  \begin{array}{l l}
 \exp\Big(-\mu_0 \,\big( pt\big)^2 \Big) & t\ll\tau\\ 
  \exp\Big(-\mu_\infty \,p^2 | t|\Big)  & t\gg\tau
 \end{array}\right.,
 \label{eq:asymp}
\end{equation}
where $\mu_0,\mu_\infty$ are non-universal constants, and $\tau$ is a typical  timescale (see App.~\ref{app:largep}). 
Let us first focus on the small time expression. It shows that the data for $C(t,p)/C(0,p)$ should collapse when plotted as a function of $pt$. Thus, it demonstrates the $z=1$ dynamical scaling exponent. This behavior is reminiscent of the effect of random sweeping in 3D turbulence, although int the case of the 1D Burgers equation,  the large scales do not dominate. Furthermore, it gives the asymptotic form of the associated scaling function, which is simply a Gaussian. This result is  compared in Fig. \ref{fig:scaling-function-IB} with the data from the numerical simulations of   \cite{Brachet2022}, using $\mu_0$ as a fitting parameter. The numerical data are accurately described by the Gaussian, as was already  argued in  \cite{Brachet2022}, at small $pt\lesssim 4$, before the negative dip which is not featured by the Gaussian, but is reproduced by the NLO solution. 
Let us now turn to the large time behavior. In the numerical data, the initial Gaussian decay is such that the scaling function rapidly reaches numerical noise  level, preventing one from resolving the large time regime and accessing the crossover at large time. However, one can notice that the quality of the collapse deteriorates at large $pt\gtrsim 4$, which  signals a change of behavior, as expected from the theoretical prediction \eqref{eq:asymp} of a $p^2 t$ scaling at large time. This would require a better resolution to be further investigated.

\sect{Conclusion.}

We have shown that the 1D KPZ equation, although exactly solvable, still reveals unforseen features, as we demonstrated the emergence of a new scaling $z=1$. This scaling  arises from the inviscid Burgers fixed point which, although unstable in the IR, controls the UV behavior of the correlation function when the initial non-linearity is large enough. We established  this scaling by numerically solving the FRG flow equations within the NLO approximation, and by obtaining the exact asymptotic form of the correlation function in the limit of large wavenumber. 

 The probability distributions of the height fluctuations are not easily accessible within the FRG formalism. The distributions in the inviscid limit have been studied numerically in \cite{Brachet2022,Rodriguez2022} which showed that they are non-Gaussian, but differ from the Tracy-Widom distributions expected at the KPZ fixed point. We hope our findings will trigger new works to obtain exact results on the distributions  at the new fixed point. They also opens up an uncharted territory which is the UV, or large non-linearity, scaling behavior of the KPZ equation in higher dimensions. 

\begin{acknowledgements}
The authors  thank B. Delamotte and N. Wschebor for enlightening discussions. CF anf FV contributed equally to this work. LC acknowledges support from the French ANR through the project NeqFluids (grant ANR-18-CE92-0019) and support from Institut Universitaire de France (IUF). FV acknowledges support from QuantForm-UGA (grant ANR-21-CMAQ-003).
\end{acknowledgements}

\begin{widetext}
\appendix
\section{General FRG framework for the KPZ and Burgers equation}
\label{app:FRG}

We study the KPZ field theory, which can be obtained from the Martin-Siggia-Rose Janssen de Dominicis formalism \cite{Martin73,Janssen76,Dominicis76,Langouche79}
\begin{align}
{\cal Z}[{\cal J}] &= \int {\cal D}h {\cal D}\tilde{h} \; e^{-{\cal S}_{\rm KPZ}[\varphi]+\int_{t,\vx} {\cal J}\cdot{\cal \varphi}} \,\label{eq:actionKPZapp}\\
{\cal S}_{\rm KPZ}[\varphi] &= \int_{t,\vx}\Big\{\tilde{h}\Big[\p_t h -\nu \nabla^2 h - \dfrac \lambda 2 \big(\vnabla h\big)^2   \Big] - D\tilde{h}^2\Big\}\, ,
\nonumber
\end{align}
where $\varphi =(h,\tilde{h})$, ${\cal J} =(J,\tilde{J})$ are the sources, and $\int_{t,\vx}\equiv \int dtd^d\vx$. Note that the derivation of the path integral formulation involves a functional determinant. 
We choose It$\bar{\rm o}$'s convention for the time discretization such that this determinant is independent of the fields and can be absorbed in the functional measure ${\cal D}\varphi$ \cite{Tauber2014}.

The exact flow equation for the effective average action $\Gamma_\kappa$ is given by 
\begin{equation}
 \partial_\kappa \Gamma_\kappa = \dfrac 1 2 {\rm Tr}\int\, \p_\kappa{\cal R}_\kappa \cdot  G_\kappa\, ,\qquad G_\kappa\equiv \Big[\Gamma_\kappa^{(2)} + {\cal R}_\kappa\Big]^{-1}\, ,
 \label{eq:dsGamk}
 \end{equation}
 where $G_\kappa$, ${\cal R}_\kappa$ and $\Gamma_\kappa^{(2)}$ are $2\times 2$ matrices in the field space $\Psi = (\psi,\tilde{\psi})$. The scalar fields $\psi$ and $\tilde\psi$ can represent the average values of the height field $\psi=\langle h\rangle$ and associated response field $\tilde\psi=\langle \tilde h\rangle$ for the KPZ equation in any dimension, or the velocity field $\psi=\langle v\rangle$ and 
 associated response  field $\tilde\psi=\langle \tilde v\rangle$ for the Burgers equation in one dimension.
 An appropriate choice for the cutoff matrix to study these equations is, in Fourier space,  
 \begin{equation}
  {\cal R}_\kappa(\omega,\vq) = \left(\begin{array}{c c} 
  0 &   {\cal M}_\kappa(\vq) \\ {\cal M}_\kappa(\vq) & {\cal N}_\kappa(\vq) 
  \end{array}\right)\, .
 \end{equation}
One can show that this form is compatible with  causality constraints \cite{Canet2011heq}, and it satisfies the Galilean symmetry. For the KPZ equation, one can parametrize the cutoff functions as ${\cal M}_\kappa(\vq) = q^2 M_\kappa(q)$ and  ${\cal N}_\kappa(\vq)= -2  N_\kappa(q)$, where $q=|\vq\,|$.  Then, in one dimension, the additional time reversal symmetry  simply imposes $N_\kappa= \frac{D}{\nu}M_\kappa$ \cite{Canet2010, Canet2011kpz}.
Note that the regulator matrix can also be chosen $\omega$-dependent, provided it satisfies the causality and symmetry constraints. 
 
 One defines the vertex functions
 \begin{equation}
  \Gamma_\kappa^{(m,n)}(t_1,\vx_1,\cdots,t_{m+n},\vx_{n+m}) = \dfrac{\delta^{m+n} \Gamma_\kappa}{\delta \psi(t_1,\vx_1)\cdots \delta \psi(t_m,\vx_m)\delta \tilde\psi(t_{m+1},\vx_{m+1})\cdots\delta \tilde\psi(t_{m+n},\vx_{m+n})}\, .
 \end{equation}
Because of translational invariance in space and time, their Fourier transforms take the form
\begin{equation}
  \Gamma_\kappa^{(m,n)}(\omega_1,\vq_1,\cdots,\omega_{m+n},\vq_{n+m}) = (2\pi)^{d+1}\delta\left(\sum_{\ell=1}^{m+n}\omega_{\ell}\right)\delta^d\left(\sum_{\ell=1}^{m+n}\vq_{\ell}\right)\bar{\Gamma}_\kappa^{(m,n)}(\omega_1,\vq_1,\cdots,\omega_{m+n-1},\vq_{n+m-1})  \, ,
\end{equation}
where the last frequency and momentum of $\bar{\Gamma}_\kappa^{(m,n)}$ are implicit as they are fixed by the conservation of the total frequency and momentum.
The Hessian of $\Gamma_\kappa$  corresponds in Fourier space to the matrix  
\begin{equation}
  \bar{\Gamma}_\kappa^{(2)}(\omega,\vq) = \left(\begin{array}{c c} 
  0 &  \bar{\Gamma}_\kappa^{(1,1)}(\omega,\vq) \\  \bar{\Gamma}_\kappa^{(1,1)}(-\omega,\vq) &  \bar{\Gamma}_\kappa^{(0,2)}(\omega,\vq)
  \end{array}\right)\, .
   \label{eq:defGam2}
 \end{equation}
 Note that because of causality, $\bar{\Gamma}_\kappa^{(2,0)}$ is always zero, and more generally   $\bar{\Gamma}_\kappa^{(m,0)}=0$ \cite{Canet2011heq}.
The propagator $\bar{G}_\kappa$ in Fourier space is simply the matrix inverse of   $[\bar{\Gamma}_\kappa^{(2)}+{\cal R}_\kappa]$  given by
\begin{equation}
 \bar{G}_\kappa(\omega,\vq)  
 = \left(\begin{array}{c c} 
  \bar{C}_\kappa(\omega,\vq) &  \bar{R}_\kappa(\omega,\vq) \\  \bar{R}_\kappa(-\omega,\vq) &  0
  \end{array}\right)\, ,\quad   \bar{R}_\kappa(\omega,\vq) = \dfrac{1}{\bar{\Gamma}_\kappa^{(1,1)}(-\omega,\vq)+{\cal M}_\kappa(\vq)} \, ,\quad  \bar{C}_\kappa(\omega,\vq) = -\dfrac{\bar{\Gamma}_\kappa^{(0,2)}(\omega,\vq)+{\cal N}_\kappa(\vq)}{\big|\bar{\Gamma}_\kappa^{(1,1)}(\omega,\vq)+{\cal M}_\kappa(\vq)\big|^2}\,.
  \label{eq:defG}
 \end{equation}
The function  $\bar{C}_\kappa = \langle h h\rangle_c$ (or $\langle v v\rangle_c$) is the connected correlation function and $\bar{R}_\kappa = \langle \tilde{h} h\rangle_c$  (or $\langle \tilde v v\rangle_c$)  is the connected response function.
 
The exact flow equations for the two-point vertex functions $\bar\Gamma^{(2)}_\kappa$ are obtained by taking two functional derivatives of \eqref{eq:dsGamk} with respect to the fields $\Psi_\alpha$ and $\Psi_\beta$. This yields in Fourier space
 \begin{align}
\p_s \bar{\Gamma}_{\kappa,\alpha\beta}^{(2)}(\varpi,\vp) & = \dfrac 1 2 {\rm Tr} \Bigg[\tilde{\p}_s  \int_{\omega,\vq} \bar{G}_{\kappa}(\omega,\vq) \cdot \Big\{ \bar{\Gamma}_{\kappa,\alpha\beta}^{(4)}(\varpi,\vp,-\varpi,-\vp,\omega,\vq) \nonumber\\
 &-   \bar{\Gamma}_{\kappa,\alpha}^{(3)}(\varpi,\vp,\omega,\vq)\cdot \bar{G}_\kappa(\varpi+\omega,\vp+\vq)\cdot \bar{\Gamma}_{\kappa,\beta}^{(3)}(-\varpi,-\vp,\varpi+\omega,\vp+\vq)\Big\}\, \Bigg], 
 \label{eq:dsgam2}
\end{align}
where the operator $\tilde{\p}_s$ only acts on the regulators, {\it i.e.} $\tilde{\p}_s\equiv \p_s {\cal R}_{\kappa ,ij}\frac{\p}{\p {\cal R}_{\kappa,ij}}$, and $s=\ln(\kappa/\Lambda)$ is the RG ``time''. The 3- and 4-point vertices $\bar{\Gamma}_{\kappa,\alpha}^{(3)}$ and $\bar{\Gamma}_{\kappa,\alpha\beta}^{(4)}$ are written as $2\times 2$ matrices, the remaining fields being fixed to the external ones (labeled by the indices $\alpha$ and $\beta$). 

 The different terms involved in the flow equation \eqref{eq:dsgam2} can be represented as diagrams. For this, let us represent the field $\psi$ by a line carrying an ingoing arrow and the response field $\tilde \psi$ by a line carrying an outgoing arrow. With this convention, the propagator components $R_\kappa$ and $C_\kappa$ in \eqref{eq:defG}, and the vertex functions       $\Gamma_\kappa^{(3)}$ and $\Gamma_\kappa^{(4)}$ can be represented as shown in \fref{fig:vertex}.
\begin{figure}[h]
\begin{tikzpicture}[scale=1.5] \path [draw=black,postaction={on each segment={mid arrow=black}}]
  (0,0) -- (0.5,0)
  (0,0) -- (-0.5,0);
   \node at (0,0) {\scriptsize $\bullet$};
  \node at (-1,0) {  $C_\kappa$};
  \end{tikzpicture}
\hspace{1cm}
\begin{tikzpicture}
  \path [draw=black,postaction={on each segment={mid arrow=black}}]
  (0,0) -- (0.5,0)
  (-0.5,0) -- (0,0);
\node at (-1,0) {  $R_\kappa$};
 \end{tikzpicture}

\vspace{0.6cm} 

\begin{tikzpicture} 
 \path [draw=black,postaction={on each segment={mid arrow=black}}]
  (0,0.35) -- (0.35,0)
  (0,-0.35) -- (0.35,0)
  (0.35,0) -- (0.85,0);
   \node at (-0.5,0) {   $\Gamma_\kappa^{(2,1)}$};
 \end{tikzpicture}
  \hspace{0.6cm}   
\begin{tikzpicture} 
 \path [draw=black,postaction={on each segment={mid arrow=black}}]
  (0,0) -- (0.5,0)
  (0.5,0) -- (0.85,0.35)
  (0.5,0) -- (0.85,-0.35);
  \node at (-0.5,0) {  $\Gamma_\kappa^{(1,2)}$};
 \end{tikzpicture}
 \hspace{0.6cm} 
\begin{tikzpicture} 
 \path [draw=black,postaction={on each segment={mid arrow=black}}]
  (0.5,0) -- (0,0)
  (0.5,0) -- (0.85,0.35)
  (0.5,0) -- (0.85,-0.35);
  \node at (-0.5,0) {   $\Gamma_\kappa^{(0,3)}$};
 \end{tikzpicture}
 
\vspace{0.6cm} 

\begin{tikzpicture} 
\path [draw=black,postaction={on each segment={mid arrow=black}}]
  (0.15,0.35) -- (0.5,0)
  (0.15,-0.35) -- (0.5,0)
  (0.5,0) -- (0.85,0.35)
  (0.5,0) -- (0.85,-0.35);
  \node at (-0.5,0) {   $\Gamma_\kappa^{(2,2)}$};
 \end{tikzpicture}
  \hspace{0.6cm}   
\begin{tikzpicture} 
 \path [draw=black,postaction={on each segment={mid arrow=black}}]
  (0.15,0.35) -- (0.5,0)
  (0.15,-0.35) -- (0.5,0)
  (0.,0) -- (0.5,0)
  (0.5,0) -- (1,0);
 \node at (-0.5,0) {   $\Gamma_\kappa^{(3,1)}$};
 \end{tikzpicture}
  \hspace{0.6cm}   
\begin{tikzpicture} 
 \path [draw=black,postaction={on each segment={mid arrow=black}}]
  (0.0,0.0) -- (0.5,0)
  (0.5,0.) -- (0.85,0.35)
  (0.5,0) -- (0.85,-0.35)
  (0.5,0) -- (1,0);
 \node at (-0.5,0) {   $\Gamma_\kappa^{(1,3)}$};
 \end{tikzpicture}
\hspace{0.6cm}   
\begin{tikzpicture} 
\path [draw=black,postaction={on each segment={mid arrow=black}}]
  (0.5,0) -- (0.15,0.35)
  (0.5,0) -- (0.15,-0.35)
  (0.5,0) -- (0.85,0.35)
  (0.5,0) -- (0.85,-0.35);
 \node at (-0.5,0) {  $\Gamma_\kappa^{(0,4)}$};
 \end{tikzpicture}
\caption{Diagrammatic representation of propagators and vertices, lines with ingoing arrows indicate the field $\psi$, and lines with outgoing arrows indicate the response field $\tilde\psi$. The propagator $C_\kappa$ is emphasized with a dot.}
\label{fig:vertex}
\end{figure}
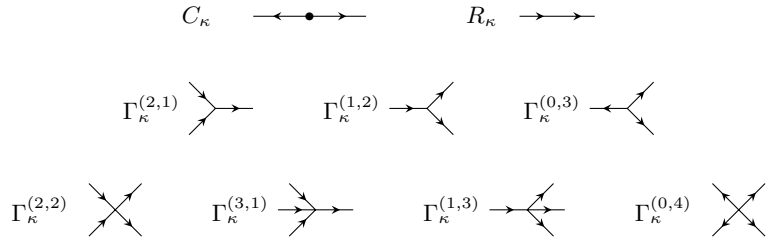

Performing the matrix products and the trace in \eqref{eq:dsgam2} for $\alpha=\beta=2$ leads to the exact flow equation for $\bar\Gamma_\kappa^{(0,2)}$, which is represented diagrammatically in \fref{fig:flowG02} (without the combinatorial factors).
\begin{figure}[h]
\begin{tikzpicture}
 \path [draw=black,postaction={on each segment={mid arrow=black}}]
  (0,0.1) -- (0.5,-0.35)
  (0,0.1) -- (-0.5,-0.35) 
  ;
  \draw (0,.5) circle(.4);
   \node at (0,-0.6) {  $(a)$};
    \node at (0,0.9) {\scriptsize $\bullet$};
  \path [draw=black,postaction={on each segment={mid arrow=black}}]
  (-0.4,0.45) -- (-0.4,0.44)
  (0.4,0.45) -- (0.4,0.44);
   \node at (-2,0.2) {   $\p_s\Gamma_\kappa^{(0,2)} = \quad \tilde\p_s\Bigg\{\;$};
    \node at (1,0.2) {  $+$};
  \path [draw=black,postaction={on each segment={mid arrow=black}}]
    (2,0.1) -- (2.5,-0.35)
    (2,0.1) -- (1.5,-0.35) 
    ;
  \draw (2,0.5) circle(.4);
   \node at (2,-0.6) {  $(b)$};
  \path [draw=black,postaction={on each segment={mid arrow=black}}]
  (1.6,0.45) -- (1.6,0.44)
  (2.4,0.54) -- (2.4,0.55);
   \node at (3,0.2) {  $-$};
    \path [draw=black,postaction={on each segment={mid arrow=black}}]
  (4.9,0.2) -- (5.5,0.2)
  (4.1,0.2) -- (3.5,0.2) 
  ;
  \draw[color=black] (4.5,0.2) circle(.4);
   \node at (4.5,-0.6) {  $(c)$};
   \node at (4.5,0.6) {\scriptsize $\bullet$};
    \node at (4.5,-0.2) {\scriptsize $\bullet$};
   \path [draw=black,postaction={on each segment={mid arrow=black}}]
  (4.85,0.41) -- (4.852,0.408)
  (4.15,0.41) -- (4.148,0.408)
  (4.15,-0.01) -- (4.148,-0.008)
  (4.85,-0.01) -- (4.852,-0.008)
  ;
   \node at (-1.5,-1.4) {  $-$};
   \path [draw=black,postaction={on each segment={mid arrow=black}}]
  (-0.4,-1.4) -- (-1,-1.4)
  (0.4,-1.4) -- (1,-1.4);
   \draw (0,-1.4) circle(.4);
     \node at (0,-2.2) {  $(d)$};
    \path [draw=black,postaction={on each segment={mid arrow=black}}]
    (0.27,-1.11) -- (0.268,-1.108)
    (0.335,-1.63) -- (0.337,-1.628)
    (-0.25,-1.71) -- (-0.248,-1.712)
   (-0.33,-1.17) -- (-0.332,-1.172)
  ;
    \node at (1.5,-1.4) {  $-$};
      \path [draw=black,postaction={on each segment={mid arrow=black}}]
     (2.6,-1.4) -- (2,-1.4)
  (3.4,-1.4) -- (4,-1.4);
   \draw (3,-1.4) circle(.4);
    \node at (3,-2.2) {  $(e)$};
     \path [draw=black,postaction={on each segment={mid arrow=black}}]
      (3.27,-1.11) -- (3.268,-1.108)
    (3.27,-1.69) -- (3.268,-1.692)
    (2.67,-1.63) -- (2.668,-1.628)
   (2.67,-1.17) -- (2.668,-1.172)
  ;
    \node at (4.5,-1.4) {  $-$};
       \path [draw=black,postaction={on each segment={mid arrow=black}}]
     (5.5,-1.4) -- (4.9,-1.4)
  (6.3,-1.4) -- (6.9,-1.4);
   \draw (5.9,-1.4) circle(.4);
    \node at (5.9,-2.2) {  $(f)$};
      \path [draw=black,postaction={on each segment={mid arrow=black}}]
      (6.23,-1.17) -- (6.232,-1.172)
    (6.17,-1.69) -- (6.168,-1.692)
    (5.57,-1.63) -- (5.568,-1.628)
   (5.57,-1.17) -- (5.568,-1.172)
  ;
   \node at (5.9,-1) {\scriptsize $\bullet$};
    \node at (7.2,-1.4) {  $\Bigg \}$};
  \end{tikzpicture}
  \caption{Diagammatic representation of the flow equation of $\bar\Gamma_\kappa^{(0,2)}$ which is given by the component $\alpha=\beta=2$ of the exact flow equation \eqref{eq:dsgam2}.}
  \label{fig:flowG02}
  \end{figure}
  
  Similarly, performing the matrix products and the trace in \eqref{eq:dsgam2} for $\alpha=1$ and $\beta=2$ leads to the exact flow equation for $\bar\Gamma_\kappa^{(1,1)}$, which is represented diagrammatically in \fref{fig:flowG11}.
  \begin{figure}[h]
\begin{tikzpicture}
 \path [draw=black,postaction={on each segment={mid arrow=black}}]
  (0,0.1) -- (0.5,-0.35)
  (-0.5,-0.35) -- (0,0.1) 
  ;
  \draw (0,.5) circle(.4);
   \node at (0,-0.6) {  $(a)$};
    \node at (0,0.9) {\scriptsize $\bullet$};
  \path [draw=black,postaction={on each segment={mid arrow=black}}]
  (-0.4,0.45) -- (-0.4,0.44)
  (0.4,0.45) -- (0.4,0.44);
   \node at (-2,0.2) {   $\p_s\Gamma_\kappa^{(1,1)} = \quad \tilde\p_s\Bigg\{ \; $};
    \node at (1,0.2) {  $+$};
  \path [draw=black,postaction={on each segment={mid arrow=black}}]
    (2,0.1) -- (2.5,-0.35)
    (1.5,-0.35) -- (2,0.1)
    ;
  \draw (2,0.5) circle(.4);
   \node at (2,-0.6) {  $(b)$};
  \path [draw=black,postaction={on each segment={mid arrow=black}}]
  (1.6,0.45) -- (1.6,0.44)
  (2.4,0.54) -- (2.4,0.55);
   \node at (3,0.2) {  $-$};
    \path [draw=black,postaction={on each segment={mid arrow=black}}]
  (4.9,0.2) -- (5.5,0.2)
  (3.5,0.2) -- (4.1,0.2)
  ;
  \draw[color=black] (4.5,0.2) circle(.4);
   \node at (4.5,-0.6) {  $(c)$};
   \node at (4.5,0.6) {\scriptsize $\bullet$};
   \path [draw=black,postaction={on each segment={mid arrow=black}}]
  (4.85,0.41) -- (4.852,0.408)
  (4.15,0.41) -- (4.148,0.408)
  (4.228,-0.088) -- (4.23,-0.09)
  (4.85,-0.01) -- (4.852,-0.008)
  ;
   \node at (6.,0.2) {  $-$};
   \path [draw=black,postaction={on each segment={mid arrow=black}}]
   (6.5,0.2) -- (7.1,0.2)
  (7.9,0.2) -- (8.5,0.2);
   \draw (7.5,0.2) circle(.4);
     \node at (7.5,-0.6) {  $(d)$};
    \path [draw=black,postaction={on each segment={mid arrow=black}}]
    (7.77,0.49) -- (7.768,0.492)
    (7.835,-0.03) -- (7.837,-0.028)
    (7.25,-0.11) -- (7.252,-0.112)
   (7.17,0.43) -- (7.168,0.428)
  ;
 \node at (9.,0.2) {$-$};
   \path [draw=black,postaction={on each segment={mid arrow=black}}]
   (9.5,0.2) -- (10.1,0.2)
  (10.9,0.2) -- (11.5,0.2);
   \draw (10.5,0.2) circle(.4);
     \node at (10.5,-0.6) {  $(e)$};
    \path [draw=black,postaction={on each segment={mid arrow=black}}]
    (10.85,0.41) -- (10.852,0.408)
    (10.835,-0.03) -- (10.837,-0.028)
    (10.25,-0.11) -- (10.252,-0.112)
   (10.25,0.510) -- (10.252,0.512)
  ;
    \node at (12.,0.2) {  $\Bigg \}$};
  \end{tikzpicture}
  \caption{Diagammatic representation of the flow equation of $\bar\Gamma_\kappa^{(1,1)}$ which is given by the component $\alpha=1$, $\beta=2$ of the exact flow equation \eqref{eq:dsgam2}.}
  \label{fig:flowG11}
  \end{figure}

\section{Flow equations for the KPZ equation within different approximations}  
\label{app:NLO}

In this section, we consider several approximations for the KPZ flow equations, which are appropriate to describe the different regimes considered in the main paper.

\subsection{Flow equations in the simplest approximation}
\label{sec:LPA}

The simplest possible approximation consists in only considering renormalized couplings, which means that the original parameters $\lambda$, $D$ and $\nu$ of the KPZ equation are promoted to $\kappa$-dependent effective parameters  $\lambda_\kappa$, $D_\kappa$ and $\nu_\kappa$.  The corresponding ansatz for $\Gamma_\kappa$ reads
\begin{equation}
\Gamma_\kappa = \int_{t,\vx} \Big\{ \tilde{\psi}\Big[\p_t \psi -\nu_\kappa \nabla^2 \psi - \dfrac{\lambda_\kappa}{2} \big(\vnabla \psi\big)^2   \Big] - D_\kappa\tilde{\psi}^2\Big\}\, .
\label{eq:anzLPA}
\end{equation}	
With this ansatz, one obtains
\begin{equation}
 \bar{\Gamma}_\kappa^{(1,1)}(\omega,\vq) = i\omega +\nu_\kappa \vq^{\,2} \,,\qquad \bar{\Gamma}_\kappa^{(0,2)}(\omega,\vq) = -2 D_\kappa\, ,
 \label{eq:gam2LPA}
\end{equation}
and the only non-vanishing vertex function is the one present in the original action
\begin{equation}
 \bar{\Gamma}_\kappa^{(2,1)}(\omega_1,\vq_1,\omega_2,\vq_2) = \lambda_\kappa \vq_1\cdot \vq_2\, .
 \label{eq:gam21}
\end{equation}
One can show that the Galilean invariance imposes that $\lambda$ is not renormalized, \ie that $\lambda_\kappa \equiv \lambda$ at all scales \cite{Frey94,Canet2011kpz}.
The flows of $\nu_\kappa$ and $D_\kappa$ can be defined as
\begin{equation}
 \partial_s D_\kappa = -\dfrac 1 2 \partial_s  \bar{\Gamma}_\kappa^{(0,2)}(\varpi,\vp)\Big|_{\varpi=0,\vp=0}\,,\qquad  \partial_s \nu_\kappa = \dfrac{1}{\vp^{\,2}} \partial_s  \bar{\Gamma}_\kappa^{(1,1)}(\varpi,\vp)\Big|_{\varpi=0,\vp=0}\,.
 \label{eq:defNu}
\end{equation}

\begin{figure}[t]
\includegraphics[width=7cm]{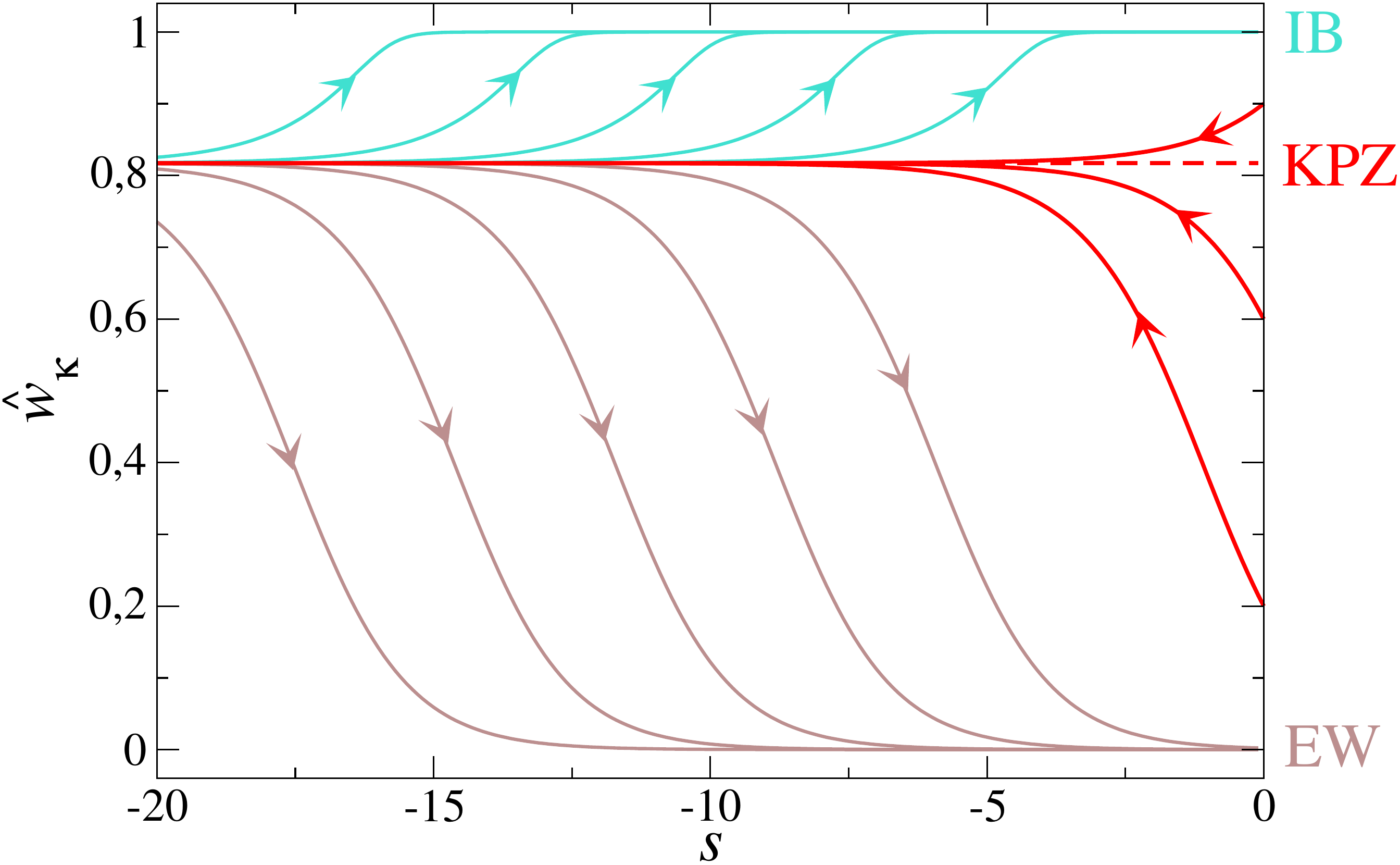}
\caption{ Flow of the coupling $\hat{w}_\kappa$ as a function of the RG time $s$ within the simple approximation of Sec.~\ref{sec:LPA}. The arrows towards decreasing $s$ indicate  IR flows, which always lead to the KPZ fixed-point for $s\to -\infty$. The arrows towards increasing $s$ indicate UV flows, which lead to either the EW or IB fixed points depending on the initial value $\hat{w}_\Lambda$ of the coupling.}
\label{fig:flow-diag}
\end{figure}
Let us now focus on one dimension (1D).
In 1D, the time reversal symmetry imposes that $D_\kappa \equiv \frac{D}{\nu}\nu_\kappa$, and their flows are equal \cite{Canet2011kpz}. Inserting \eqref{eq:defG} with \eqref{eq:gam2LPA} and \eqref{eq:gam21} in \eqref{eq:dsgam2}, one obtains from \eqref{eq:defNu}
\begin{equation}
 \p_s \nu_\kappa = 2 \,g \, \int_{\omega,q} \p_s M_\kappa(q)\, q^4 h(q)\, \dfrac{\omega^2 - \big( q^2 h_\kappa(q)\big)^2}{\Big(\omega^2 + \big(q^2 h_\kappa(q)\big)^2\Big)^3}\,, \qquad h_\kappa(q) =  \nu_\kappa + M_\kappa(q)\, ,
\end{equation}
with $g=\lambda^2 D/\nu^3$.
Carrying out the frequency integral yields
\begin{equation}
 \p_s \nu_\kappa = -g \int_{-\infty}^\infty \,\dfrac{dq}{2\pi}\, \dfrac{\p_s M_\kappa(q)}{4 \,q^2 h_\kappa(q)}\, .
\end{equation}
In order to study the fixed point, one introduces dimensionless quantities, denoted with a hat symbol. Momenta are measured in units of $\kappa$, \eg $\hat{q} = q/\kappa$ and frequencies in units of $\kappa^2 \nu_\kappa$, \eg $\hat{\omega} = \omega/(\kappa^2\nu_\kappa)$. The cutoff function can be written as
 $M_\kappa(q) = \nu_\kappa \hat{m}(y)$ with $y=q^2/\kappa^2$, and thus
\begin{equation}
 \p_s M_\kappa(q) = -\nu_\kappa \big(\eta_\kappa \hat{m}(y) + 2 y\,\hat{m}'(y)\big)\, .
\end{equation}
One obtains for the running anomalous dimension $\eta_\kappa$ 
\begin{equation}
 \eta_\kappa \equiv -\p_s \ln\nu_\kappa = -\dfrac{\hat{g}_\kappa}{8\pi} \int_0^\infty dy\, \dfrac{\eta_\kappa \hat{m}(y) + 2 y \,\hat{m}'(y)}{y^{3/2}\big(1+\hat{m}(y)\big)^2} \,,
\end{equation}
with the dimensionless coupling $\hat{g}_\kappa  = \kappa^{-1} g /\nu^2_\kappa$. One deduces the explicit expression for $\eta_\kappa$ as
\begin{equation}
 \eta_\kappa = \dfrac{\hat{g}_\kappa{\cal I}_0}{1+\hat{g}_\kappa{\cal I}_1} \, ,\quad {\cal I}_0 =  -\dfrac{1}{4\pi} \int_0^\infty dy\, \dfrac{  \hat{m}'(y)}{y^{1/2}\big(1+\hat{m}(y)\big)^2} \,,\quad {\cal I}_1= \dfrac{1}{8\pi} \int_0^\infty dy\, \dfrac{ \hat{m}(y)}{y^{3/2}\big(1+\hat{m}(y)\big)^2} \, .
\label{eq:flowetaLPA}
\end{equation}

It is clear that the two integrals ${\cal I}_0$ and ${\cal I}_1$ depend on the choice of the cutoff function $\hat{m}(y)$. For a typical choice
$\hat{m}(y) = 1/(e^y -1)$, one obtains for example ${\cal I}_0 = 1/(4\sqrt{\pi})$ and ${\cal I}_1 = (\sqrt{2}-1)/(4\sqrt{\pi})$, but they can take roughly any value for other choices, such that the value of $\eta_*$ cannot be properly determined within this approximation.
The flow equation for $\hat{g}_\kappa$   reads
\begin{equation}
 \p_s \hat{g}_\kappa  = \hat{g}_\kappa \big(-1+2\eta_\kappa\big) \,,
 \label{eq:flowg}
\end{equation}
or equivalently the one for $\hat{w}_\kappa=\hat{g}_\kappa/(1+\hat{g}_\kappa)$ is given by
\begin{equation}
 \p_s \hat{w}_\kappa = \hat{w}_\kappa \big(1-\hat{w}_\kappa\big)\big(-1+2 \eta_\kappa \big)  \,.
 \label{eq:flowwk}
\end{equation}
One can readily check that it yields the three fixed point solutions EW ($\hat{w}_*=0$), KPZ ($\eta_*=1/2$, $0<\hat{w}_*<1$) and IB ($\hat{w}_*=1$) presented in the main paper. The flow equations \eqref{eq:flowwk}, \eqref{eq:flowetaLPA}  can also be integrated numerically, the result is shown in \fref{fig:flow-diag}. Starting from any initial condition $\hat{w}_\Lambda$, the flow reaches in the IR when $\kappa\to 0$, \ie $s\to -\infty$, the KPZ fixed point $\hat{w}_*\equiv \hat{w}_{\rm \tiny KPZ}$, which is thus fully attractive and stable. To evidence the other fixed points, which are IR unstable, one can run the flow backwards, towards the UV. Depending on the initial condition $\hat{w}_{s_0}$ at some large $s_0\lesssim -20$, the flow either reaches the EW fixed point with $\hat{w}_*=0$ for any $\hat{w}_{s_0}<  \hat{w}_{\rm \tiny KPZ}$, or it leads to the IB fixed point with $\hat{w}_*=1$  for any  $\hat{w}_{s_0}> \hat{w}_{\rm \tiny KPZ}$.
This thus confirms the existence of the IB fixed point.

\subsection{Flow equations in the next-to-leading order approximation}

Whereas the simple approximation presented  in the previous section suffices to demonstrate the existence of the IB fixed point, it does not allow to obtain  quantitative results for the statistical properties of the system at this fixed point, in particular the critical exponent $z_{\tiny \rm IB}$ and the two-point correlation function. 
Let us present a more refined approximation, referred to as next-to-leading order (NLO), which was introduced in \cite{Kloss2012}, and which will be used in the numerical procedure detailed in Sec.~\ref{sec:num}. The NLO approximation was shown to yield accurate results for the scaling function at the KPZ fixed point in one dimension, where it can be compared with exact results  \cite{Kloss2012}. Moreover, this approximation can be implemented in any dimension and was used to obtain the scaling functions at the KPZ fixed point in dimensions two and three. These results were subsequently confirmed with great precision in numerical simulations \cite{Halpin-Healy13,Halpin-Healy13Err,Deligiannis2022}.
We refer the reader to \cite{Kloss2012} for a detailed presentation and justification of this approximation. Let us mention that in 1D, a yet finer approximation, called SO (second order) was devised in \cite{Canet2011kpz} and was shown to reproduce with even greater accuracy, down to very fine details of the tails, the exact result from \cite{Praehofer04} for the KPZ scaling function. 


The NLO approximation corresponds to the following ansatz
\begin{equation}
\Gamma_\kappa = \int_{t,\vx} \Big\{ \tilde{\psi}\Big[\p_t \psi -f^\nu_\kappa(\p_t,\nabla) \nabla^2 \psi - \dfrac{\lambda_\kappa}{2} \big(\vnabla \psi\big)^2   \Big] - f^D_\kappa(\p_t,\nabla)\tilde{\psi}^2\Big\}\, .
\label{eq:anzNLO}
\end{equation}	
 The Galilean symmetry yields again that $\lambda_\kappa \equiv \lambda$ is not renormalized. From the NLO ansatz, one obtains in Fourier space
\begin{equation}
 \bar{\Gamma}_\kappa^{(1,1)}(\omega,\vq) = i\omega +f^\nu_\kappa(\omega,\vq\,) \vq^{\,2} \,,\qquad \bar{\Gamma}_\kappa^{(0,2)}(\omega,\vq\,) = -2 f^D_\kappa(\omega,\vq)\, ,
 \label{eq:gam2NLO}
\end{equation}
and the only non-vanishing vertex function is still $\Gamma^{(2,1)}_\kappa$ given by \eqref{eq:gam21}. 

At NLO, the only diagram contributing to the flow equation of  $\bar\Gamma_\kappa^{(0,2)}$  is thus $(c)$
\begin{center}
\begin{tikzpicture}
 \node at (-1.8,0.2) { $\p_s \bar\Gamma_\kappa^{(0,2)} = \; -\; \tilde\p_s \; $};
    \path [draw=black,postaction={on each segment={mid arrow=black}}]
  (0.9,0.2) -- (1.5,0.2)
  (0.1,0.2) -- (-0.5,0.2) 
  ;
  \draw[color=black] (0.5,0.2) circle(.4);
   \node at (0.5,0.6) {\scriptsize $\bullet$};
    \node at (0.5,-0.2) {\scriptsize $\bullet$};
   \path [draw=black,postaction={on each segment={mid arrow=black}}]
  (0.85,0.41) -- (0.852,0.408)
  (0.15,0.41) -- (0.148,0.408)
  (0.15,-0.01) -- (0.148,-0.008)
  (0.85,-0.01) -- (0.852,-0.008)
  ;
\end{tikzpicture}
\end{center}
In 1D,  the time reversal symmetry further imposes $f^D_\kappa=\frac{D}{\nu}f^\nu_\kappa\equiv \frac{D}{\nu} f_\kappa$ \cite{Canet2011kpz,Kloss2012}. 
Inserting \eqref{eq:defG} with \eqref{eq:gam2NLO} and \eqref{eq:gam21} in \eqref{eq:dsgam2}, one obtains 
\begin{align}
 \p_s f_\kappa(\varpi, p) \equiv {\cal I}_\kappa(\varpi,p) &=  2 \,g\, \int_{\omega,q}   \partial_s M_\kappa(q) \,q^2 \frac{ (p+q)^2 h_\kappa(\varpi+\omega, p+q)\big[\omega^2 - \big(q^2 h_\kappa(\omega,q)\big)^2\Big]}{\Big(\omega^2 + \big(q^2 h_\kappa(\omega,q) \big)^2\Big)^2 \Big((\varpi+\omega)^2 + \big( (p+q)^2 h_\kappa(\varpi+\omega, p+q)\big)^2 \Big)^2}\,\nonumber\\  h_\kappa(\omega,q) &= f_\kappa(\omega,q) + M_\kappa(q) \,.
 \end{align}
 One introduces the dimensionless function $\hat{f}_\kappa(\hat{\varpi}, \hat{p}) = {f}_\kappa(\varpi=\nu_\kappa \kappa^2\hat{\varpi}, p=\kappa\hat{p})/\nu_\kappa$. Its flow equation reads 
 \begin{equation}
\p_s \hat{f}_\kappa(\hat{\varpi}, \hat{p}) = \big(\eta_\kappa + \hat{p}\p_{\hat p} + (2-\eta_\kappa) \hat{\varpi} \p_{\hat\varpi} \big) \hat{f}_\kappa(\hat{\varpi}, \hat{p})
+ \hat{{\cal I}}_\kappa(\hat{\varpi}, \hat{p})\, ,
\label{eq:pshatf}
 \end{equation}
 with $\hat{{\cal I}}_\kappa(\hat{\varpi}, \hat{p})={\cal I}_\kappa(\varpi,p)/\nu_\kappa$.
 The running anomalous dimension $\eta_\kappa$ can be defined  through the normalization condition $f_\kappa(\varpi=0,p=0) = \nu_\kappa$, or equivalently $\hat{f}_\kappa(\hat{\varpi}=0, \hat{p}=0)=1$. Its flow equation is thus obtained by 
setting  $\p_s \hat{f}_\kappa(\hat{\varpi}=0, \hat{p}=0)= 0$ in \eqref{eq:pshatf}. 
In fact, the NLO approximation includes a further simplification, which stems from neglecting the frequency dependence of the function $\hat{h}_\kappa$ within the integral. This is justified because the variation of $\hat{h}$ with the frequency is small in general, and is discussed in more details in \cite{Kloss2012}. This simplification amounts to replacing within the integrals $\hat{h}_\kappa(\Omega,Q)$ with $\hat{k}_\kappa(Q)\equiv \hat{h}_\kappa(0,Q)$ for any configuration $(\Omega,Q)$. This replacement allows one to carry out analytically the frequency integral. This leads to the  explicit expression for $\eta_\kappa$
 \begin{equation}
 \eta_\kappa = \dfrac{\hat{g}_\kappa{\cal I}_0}{1+\hat{g}_\kappa{\cal I}_1} \, ,\quad {\cal I}_0 =  -\dfrac{1}{4\pi} \int_0^\infty dy\, \dfrac{  \hat{m}'(y)}{y^{1/2}\big(\hat{k}_\kappa(y)\big)^2} \,,\quad {\cal I}_1= \dfrac{1}{8\pi} \int_0^\infty dy\, \dfrac{ \hat{m}(y)}{y^{3/2}\big(\hat{k}_\kappa(y)\big)^2} \,,\quad \hat{k}_\kappa(y) = \hat{f}_\kappa(0,y) +\hat{m}_\kappa(y)\, .
 \label{eq:eta}
\end{equation}
 These flow equations can be solved numerically, as we now describe.

\subsection{Numerical integration of the NLO flow equations}
\label{sec:num}

 In order to solve the NLO flow equations \eqref{eq:pshatf},  \eqref{eq:flowg}
 and \eqref{eq:eta}, we use the numerical scheme  introduced in Ref. \cite{Kloss2012}.
  The integration in the RG time  $s = \ln(\kappa/\Lambda)$ is performed with an explicit Euler time stepping with $ds=10^{-4}$, starting at $s=0$ ($\kappa=\Lambda$)  from the initial condition $\hat{f}_\Lambda(\hat{\varpi},\hat{p})\equiv 1$ with a fixed value of $\hat g_\Lambda$, until a fixed point is reached (typically for $s\lesssim -20$).
The dimensionless frequency  $\hat{\varpi}$ and momentum $\hat{p}$ are discretized on a logarithmic grid of $N_\varpi \times N_p$ points in $[0,\hat{\varpi}_{\rm max}]\times[0,\hat{p}_{\rm max}]$, 
 and the function $\hat{f}_\kappa(\hat{\varpi},\hat{p})$ is represented as  a $N_\varpi\times N_p$ matrix. 
A third order polynomial spline is used to compute $\hat{f}_\kappa(\hat{\varpi}+\hat{\omega},\hat{p}+\hat{q})$ for momenta and frequencies which do not fall on the tabulated grid points. This spline is also used to compute the derivatives in the linear part  of \eqref{eq:pshatf}.

To compute the integrals $\hat{\cal I}_\kappa(\hat{\varpi},\hat{p})$ in \eqref{eq:pshatf}, 
 the NLO simplification for the frequency dependence is used, which consists in 
 the replacement $\hat{f}_\kappa(\hat{\Omega},\hat{Q}) \rightarrow \hat{f}_\kappa(0,\hat{Q})$ for all configurations 
  $\hat \Omega$ and $\hat Q$ inside the integrands. This is exploited to perform the 
 integration over the internal frequency $\hat{\omega}$ analytically. The integral over the internal momentum $\hat{q}$ is computed using 
 a Gauss-Legendre quadrature. Because of the insertion of the scale derivative of the regulator,  the remaining integrand, after frequency integration,  is a smooth 
 function which is exponentially suppressed for $\hat{q} \gg 1$. 
 Followingly, the integral can be cut at $\hat{q}\simeq 6$ without loss of precision.
We now present the results.

\subsubsection{The KPZ fixed point} 

For any initial value of $\hat g_\Lambda$, the flow reaches a fixed point, where $\hat{g}_\kappa\to \hat{g}_*$ and the function $\hat{f}_\kappa(\hat{\varpi},\hat{p})$ converges to a fixed form $\hat{f}_*(\hat{\varpi},\hat{p})$.  
 A typical evolution with the RG time of $\hat{f}_\kappa(\hat{\varpi},\hat{p})$ at a fixed $\hat{\varpi}$  is shown in \fref{fig:NLO-KPZ}.
 One deduces the correlation function as 
 \begin{equation}
 \bar{C}(\hat\varpi,\hat p) =\dfrac{2 {\hat{f}_*}(\hat\varpi, \hat p)}{\hat\varpi^2 + \hat p^4 {\hat{f}_*}^2(\hat\varpi, \hat p)} \,,
 \end{equation}
 which is also represented in \fref{fig:NLO-KPZ}.  To evidence the dynamical scaling exponent, one can compute the half-frequency, defined as
 \begin{equation}
 {C}(\hat{\varpi}_{1/2}(p),\hat{p}) = \dfrac{1}{2}  {C}(0,\hat{p})\,,
 \end{equation}
 which is displayed in \fref{fig:NLO-KPZ}. It follows the scaling $\hat{\varpi}_{1/2}\sim \hat{p}^{3/2}$
 as expected for the KPZ fixed point.
  We then compute the inverse Fourier transform in time of $\bar{C}(\hat\varpi, \hat p)$ defined, exploiting parity in $\hat\varpi$, as
 \begin{equation} 
 {C}(\hat t,\hat p) = \int_{0}^{\infty} \dfrac{d\hat \varpi}{\pi}\,  \bar{C}(\hat\varpi,\hat p)\, \cos(\hat{\varpi} \hat t\,)\, .
 \end{equation}
 The data for ${C}(\hat t, \hat p)/C(0,\hat p)$ all collapse into a single one-dimensional curve, the universal KPZ scaling function, when plotted as a function of the scaling variable $\hat p \hat t^{2/3}$. This scaling function is compared in \fref{fig:NLO-KPZ} with the exact curve from \cite{Praehofer04}. The result for the more precise SO approximation from \cite{Canet2011kpz} is also shown for reference. Both the NLO and SO curves are in precise agreement with the exact result. The inset magnifies the behavior of the tail of the function, which shows that the SO approximation even reproduces down to very small amplitudes the  details of this tail, which follows a stretched exponential with superimposed oscillations on the scale $p^{3/2}$. 
 \begin{figure}[t]
 \tikz{
 \node at (-1,0) {\includegraphics[width=8cm]{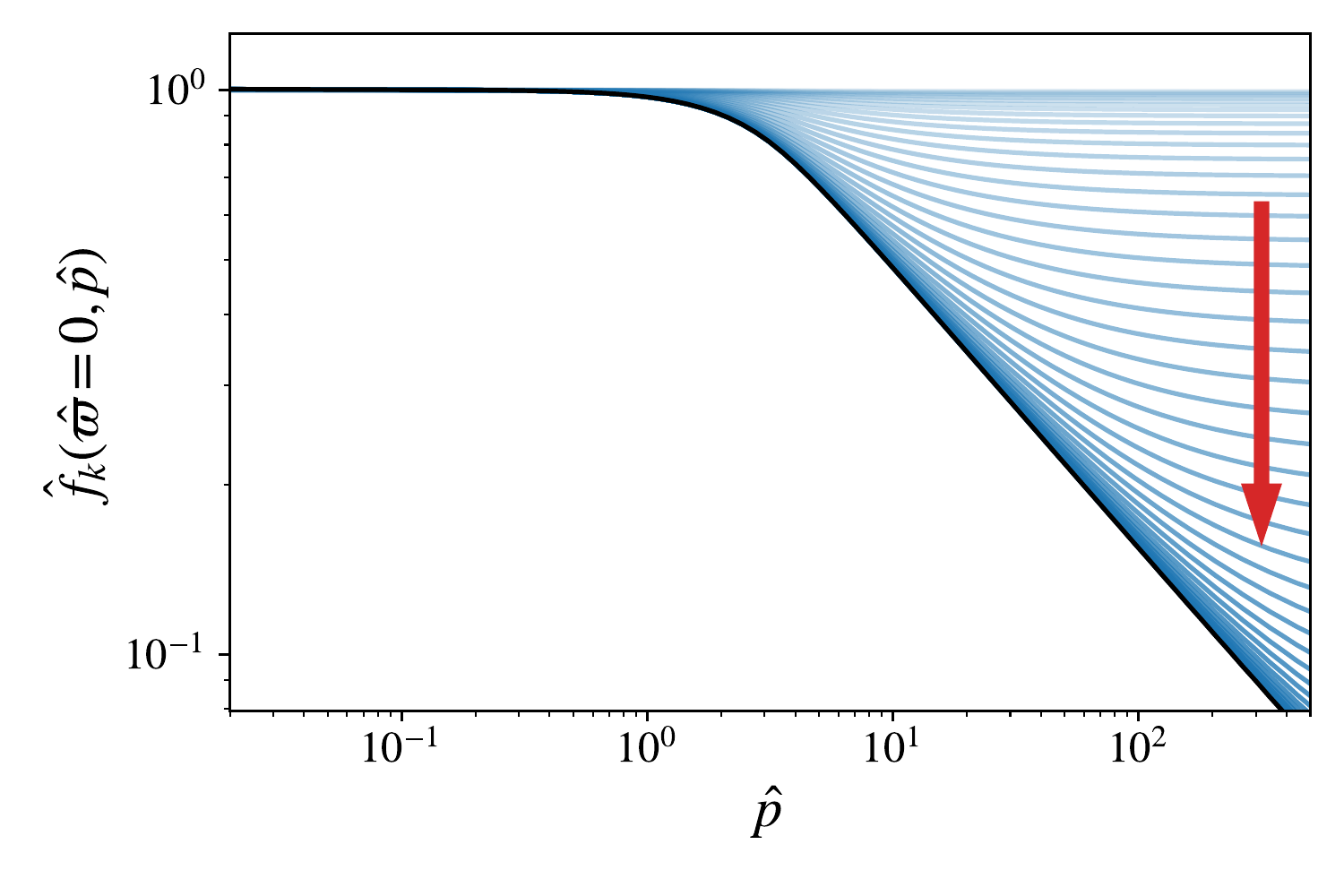}};
 \node at (8,0.5) {\includegraphics[width=8cm]{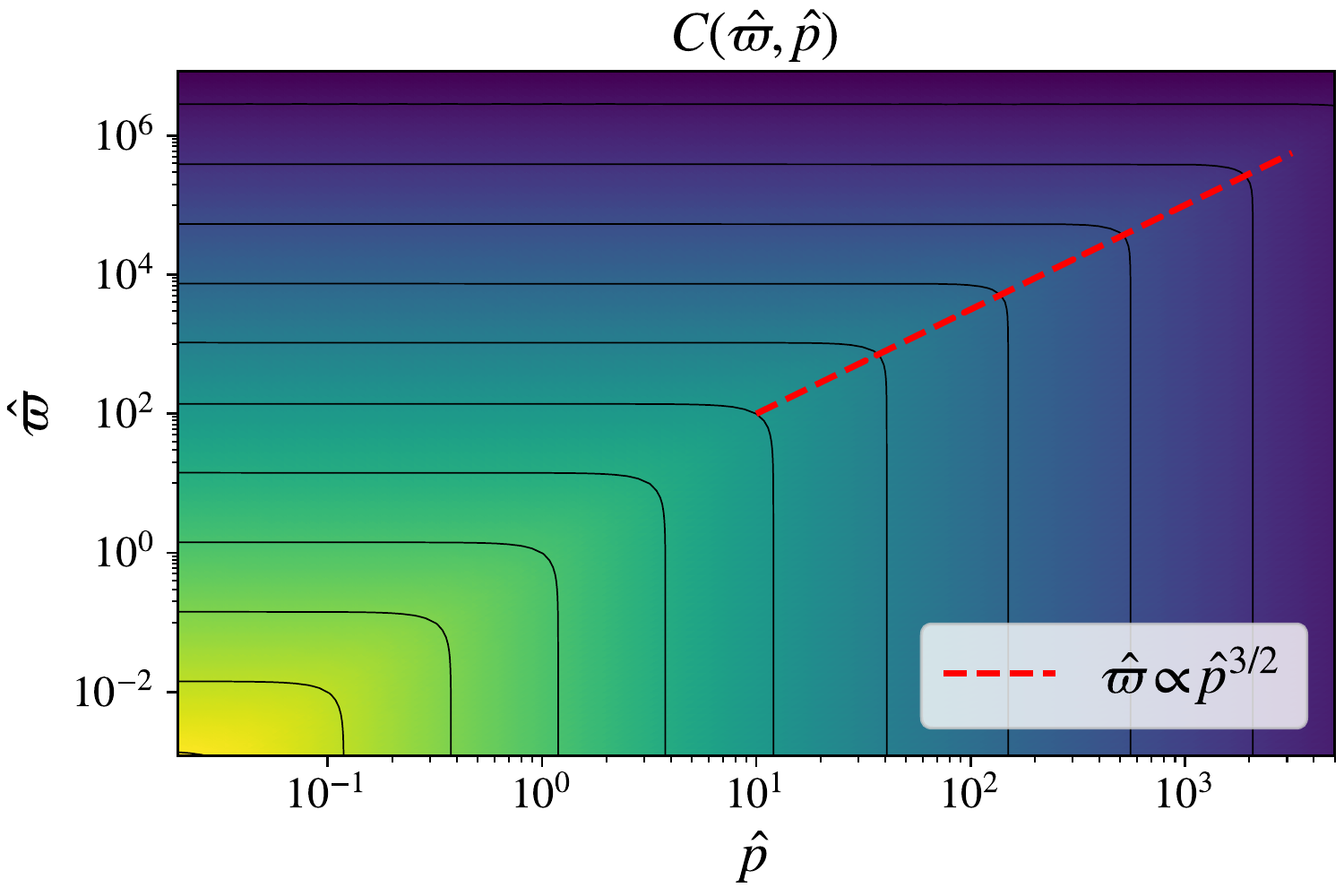}};
 \node at (-1,-5.5) {\includegraphics[width=8cm]{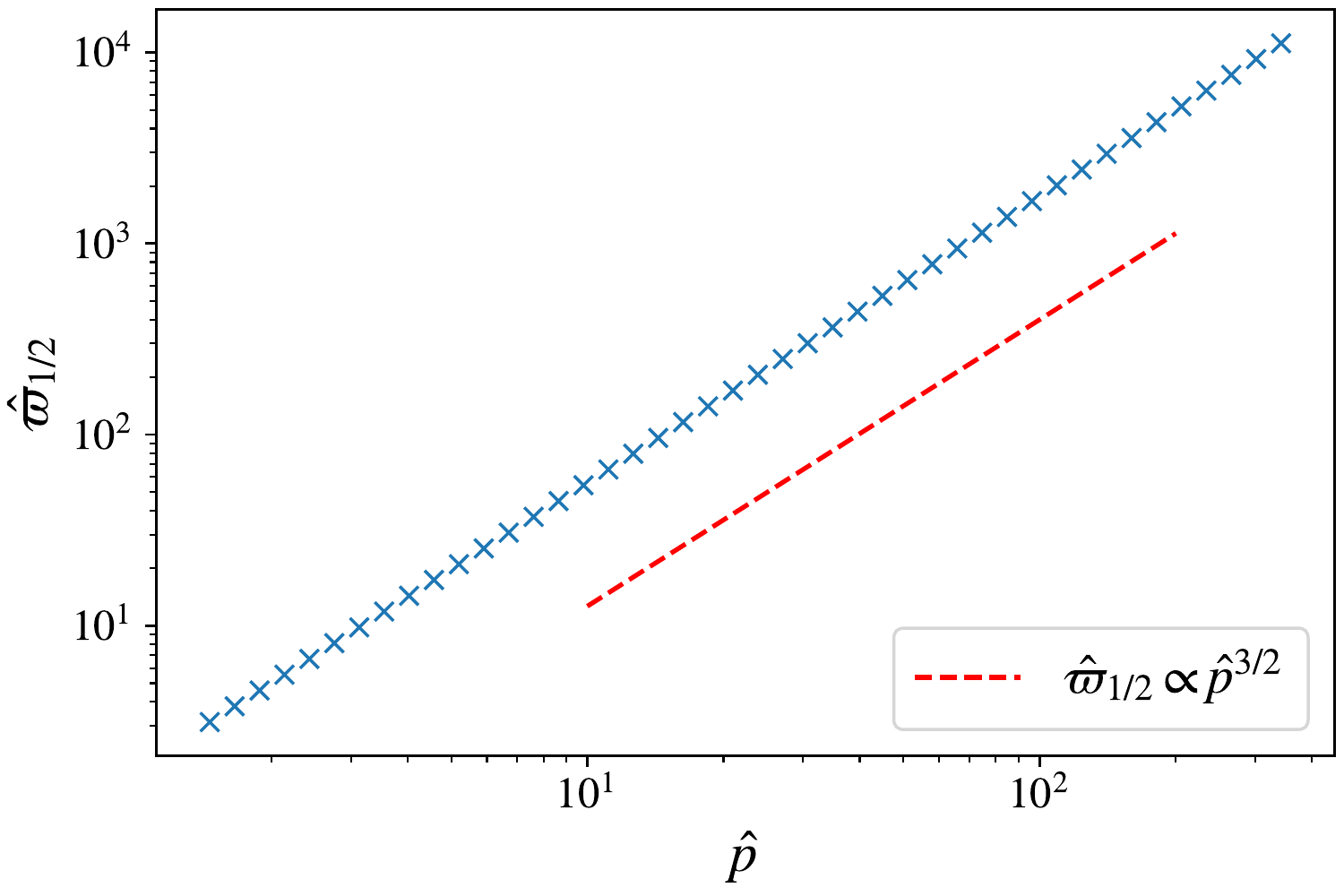}};
 \node at (8,-5.2) {\includegraphics[width=8cm]{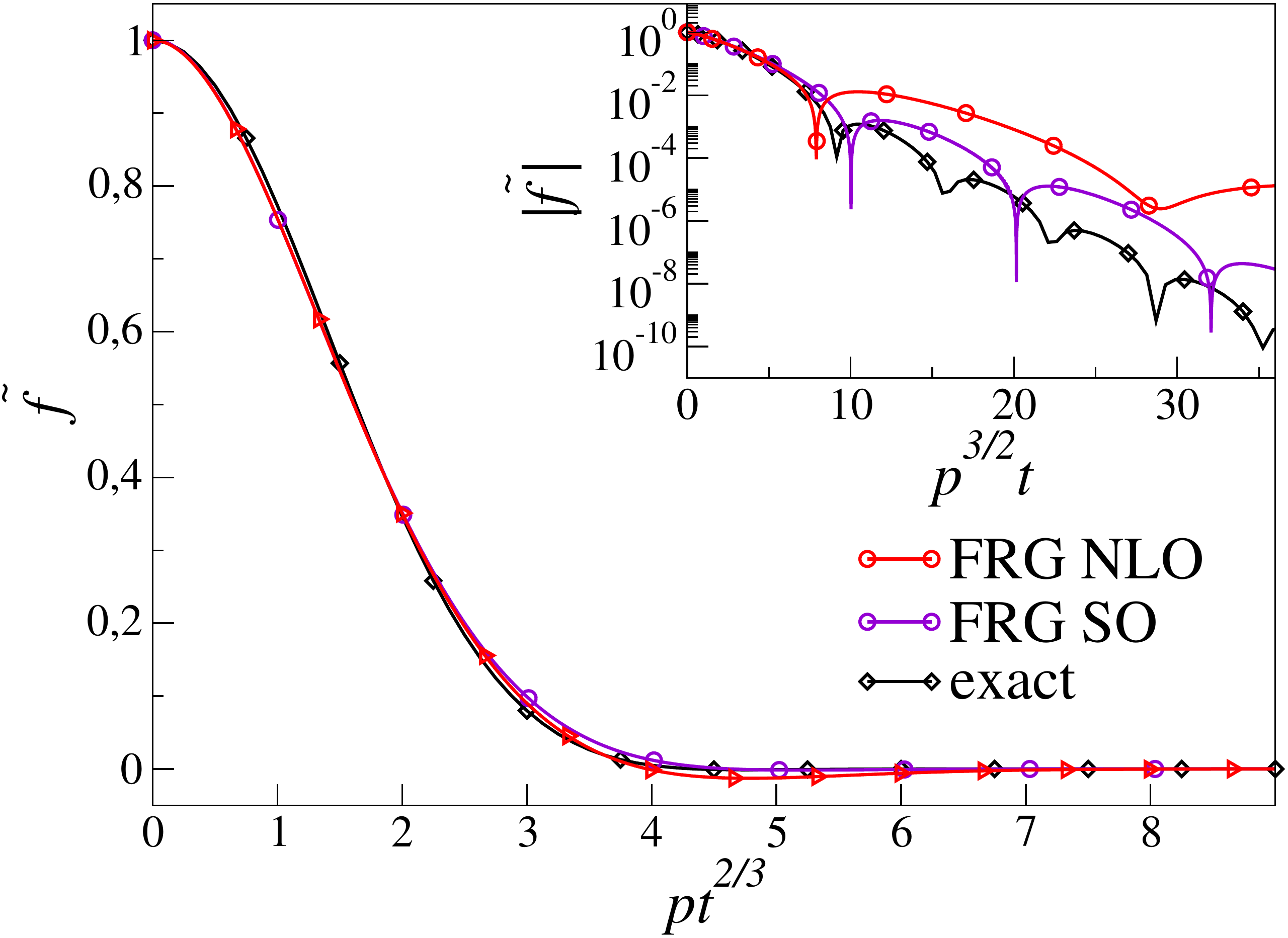}};
 } 
  \caption{({\it Top left panel}) Evolution of $\hat f_\kappa(\hat \varpi=0,\hat p)$ with the RG flow within the NLO approximation, from the constant initial condition $\hat f_\Lambda(\hat \varpi=0,\hat p)=1$ to the IR  fixed point which behaves at large $\hat p$ as $\hat f_*(\hat \varpi=0,\hat p)\sim \hat p^{-1/2}$. ({\it Top right panel}) Correlation function $C(\hat\varpi,\hat  p)$, and ({\it bottom left panel}) half-frequency  $\hat\varpi_{1/2}$ as a function of $\hat p$, which both show the KPZ dynamical exponent $\varpi\sim p^{z}$ with $z=3/2$.  ({\it Bottom right panel}) Collapse of the correlation function onto the KPZ scaling function, from FRG within the NLO approximation (this work) and from the more refined SO approximation (from \cite{Canet2011kpz}), compared with the exact result from \cite{Praehofer04}.}
  \label{fig:NLO-KPZ}
 \end{figure}
 
The initial value of  $\hat g_\Lambda$ has no influence on the fixed point reached at the end of the flow, in the IR, which is always the KPZ one. However, the beginning of the flow, in the UV, is sensitive to this value. For small values 
$\hat g_\Lambda \ll \hat{g}_*$, one expects the UV flow to be controlled by the EW fixed point, whereas for large values $\hat g_\Lambda \gg \hat{g}_*$, one expects it to be controlled by the IB fixed point. Let us determine the corresponding statistical properties.

\subsubsection{The EW fixed point} 

 \begin{figure}[t]
 \tikz{
 \node at (-1,0) {\includegraphics[width=8cm]{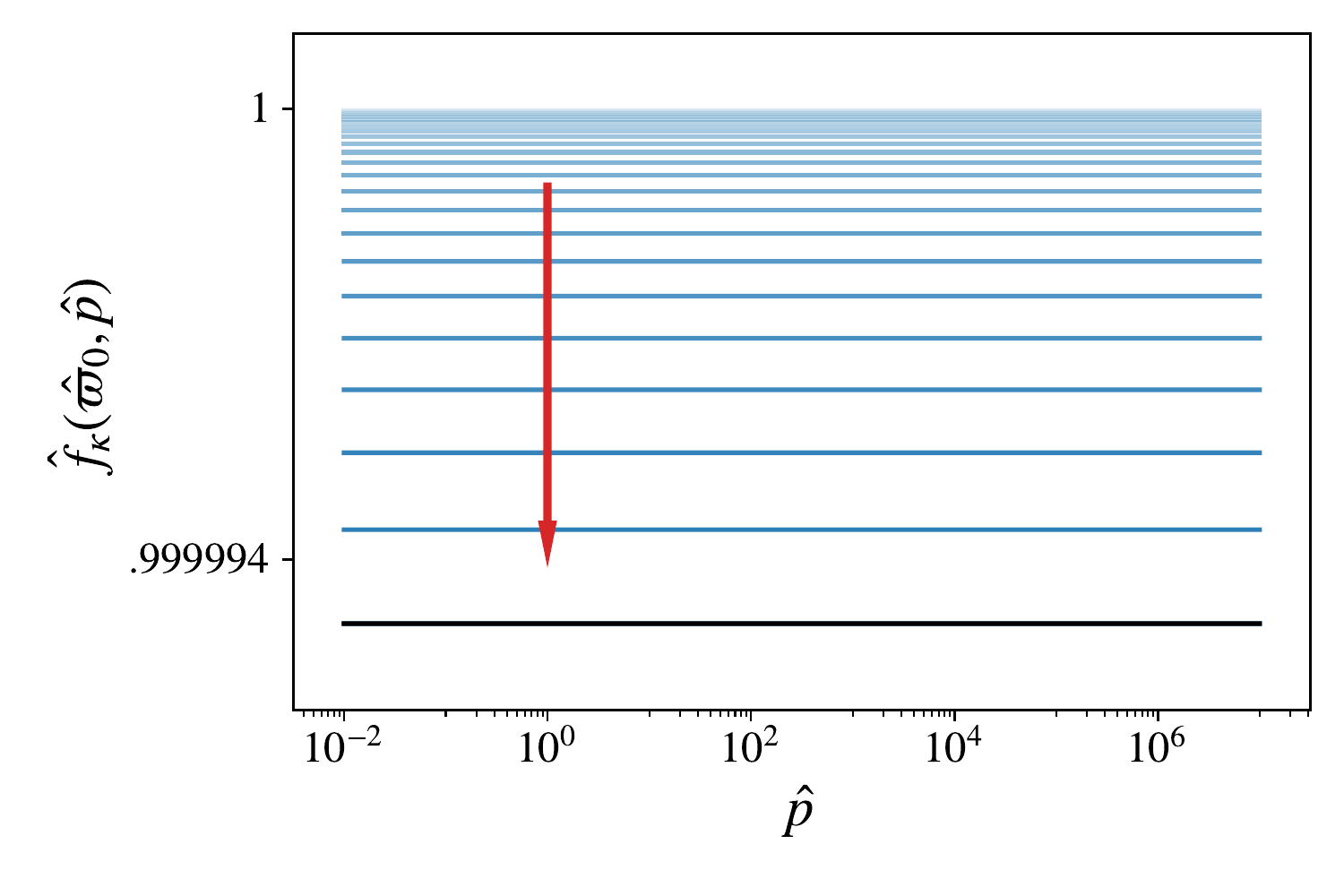}};
 \node at (8,0.5) {\includegraphics[width=8cm]{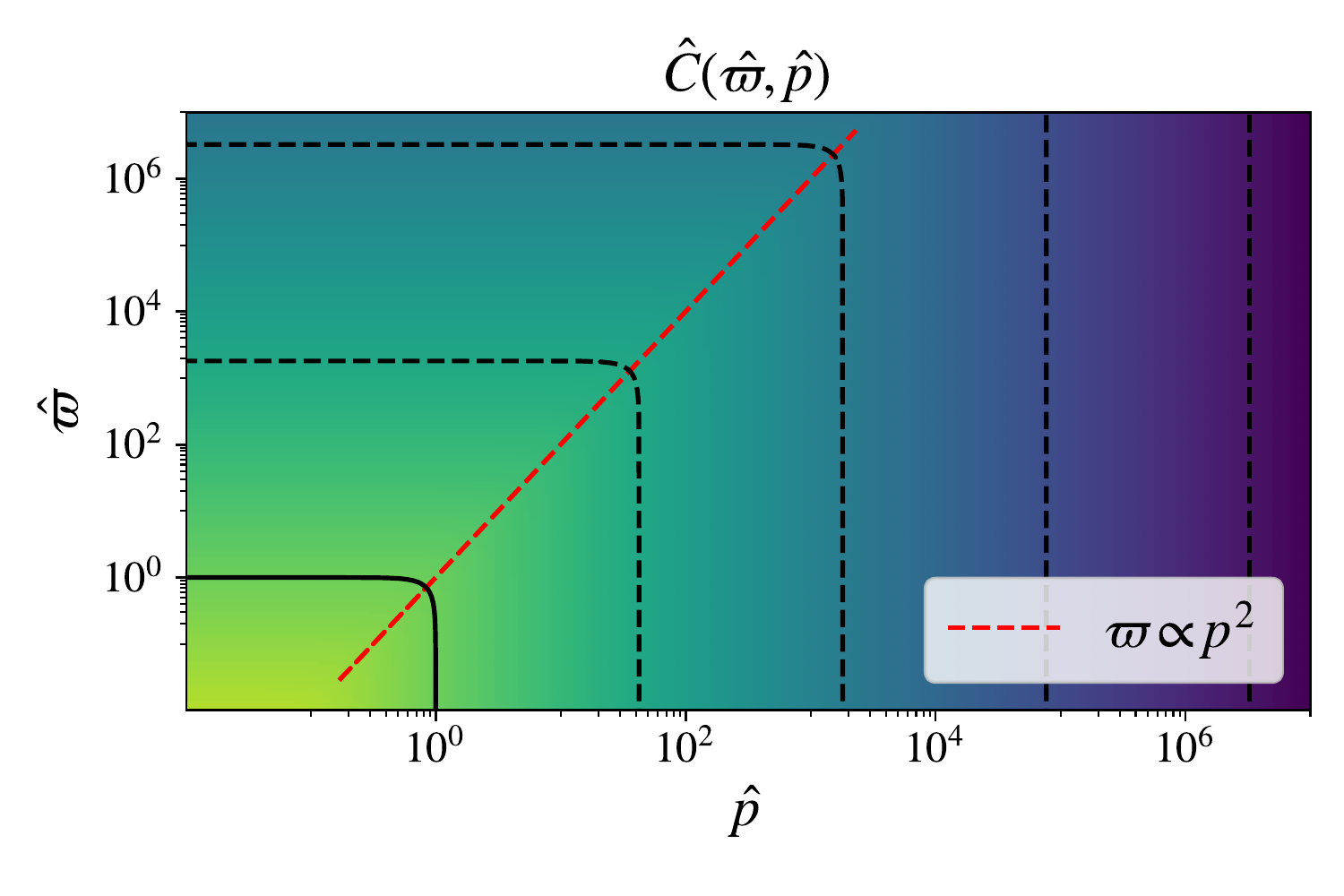}};
 \node at (-1,-5.5) {\includegraphics[width=8cm]{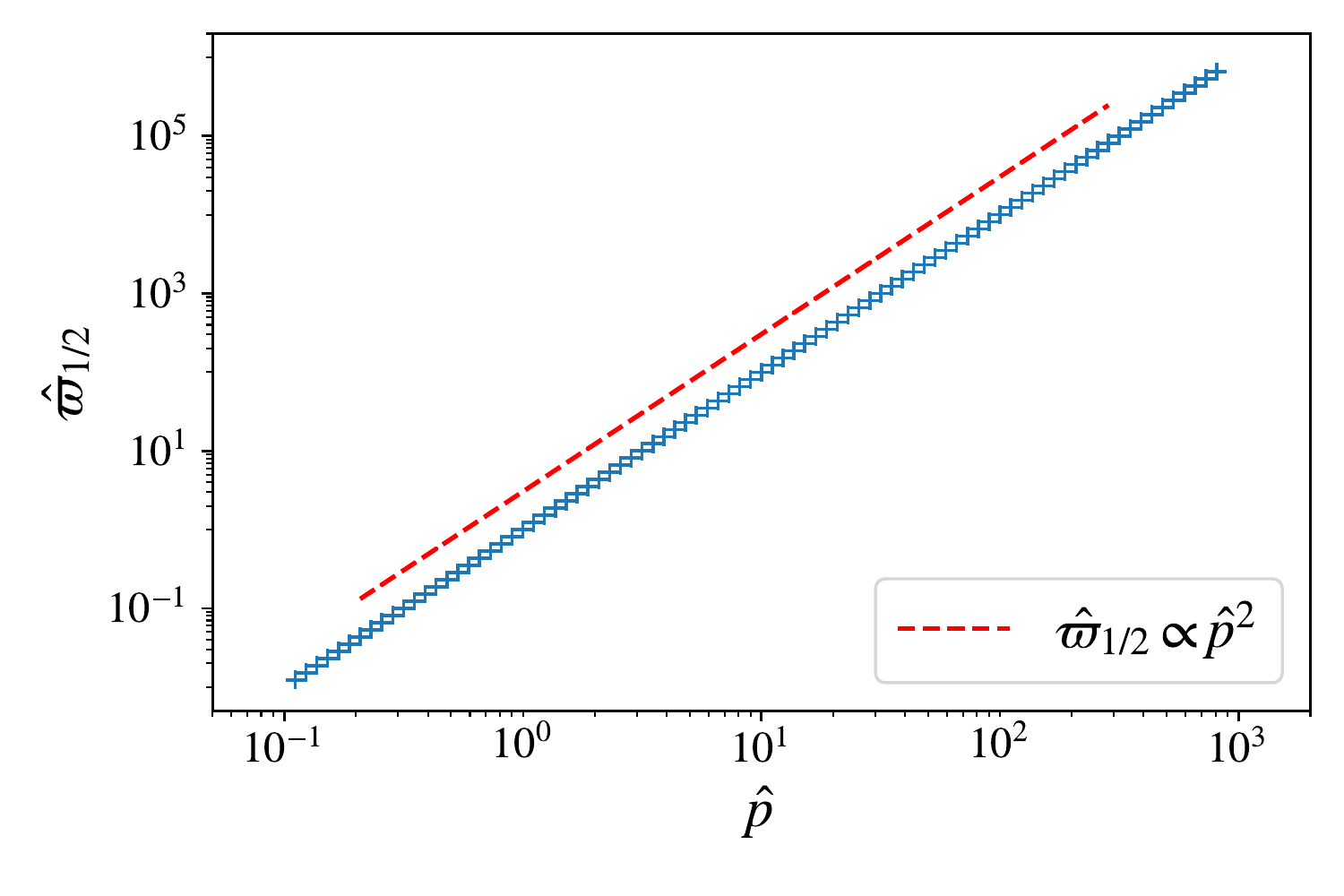}};
 \node at (8,-5) {\includegraphics[width=8cm]{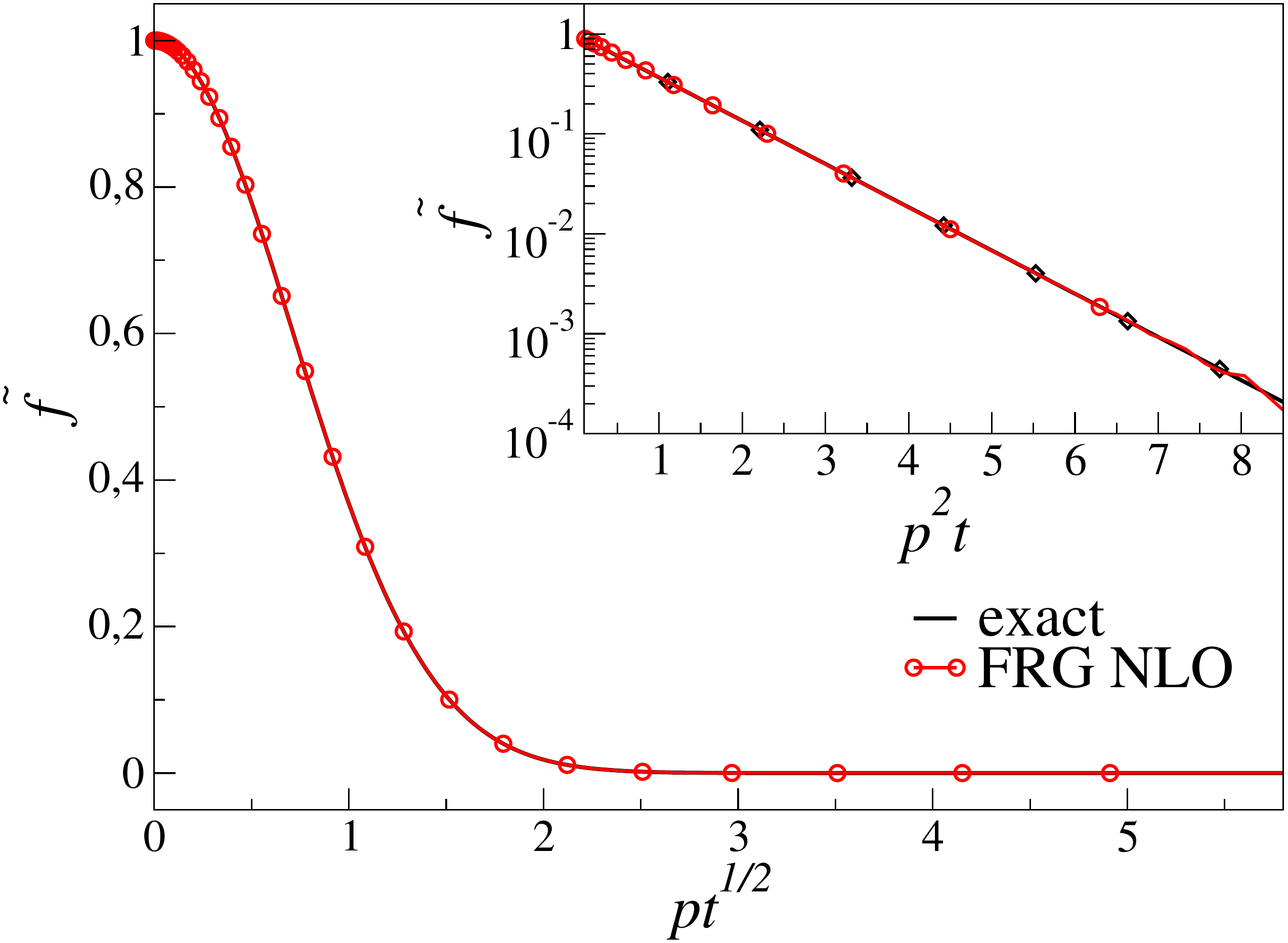}};
 } 
  \caption{({\it Top left panel}) Evolution of $\hat f_\kappa(\hat \varpi=0,\hat p)$ during the initial UV flow from a small value of $\hat g_\Lambda =10^{-6}$. ({\it Top right panel}) Correlation function $C(\hat \varpi, \hat p)$, and  ({\it bottom left panel})  half-frequency  $\hat\varpi_{1/2}$ as a function of $\hat p$, which both show the EW dynamical exponent $\varpi\sim p^{z}$ with $z=2$.  ({\it Bottom right panel}) Collapse of the correlation function onto the EW scaling function, compared with the exact result \eqref{eq:ftilde-EW-exact}.}
  \label{fig:NLO-EW}
 \end{figure}
Let us first emphasize that, since the EW equation is linear, the correlation function can be calculated exactly, without resorting to RG. One obtains in Fourier space (in dimensionful variables)
\begin{equation}
 \bar{C}(\varpi,  p) = \dfrac{2D}{\varpi^2 + \nu^2 p^4} \equiv \dfrac{D}{\nu ^2  p^4} \mathring{F}\left(\dfrac{\varpi}{\nu p^2}\right)\, ,\quad \mathring{F}(x) = \dfrac{2}{x^2 + 1}\, .
\end{equation}
The scaling form for $C(t,p)$ reads
\begin{equation}
 C(t,p) =  \int_{0}^\infty \dfrac{d\varpi}{\pi}\,\bar{C}(\varpi, p) \,\cos(\varpi t)\equiv \dfrac{D}{\nu p^2}f(\nu p t^{1/2})\, ,\quad f(x) = \int_0^\infty \dfrac{d\tau}{\pi} \, \mathring{F}(\tau)\,\cos(\tau x^2) = e^{-x^2}\, ,
  \label{eq:ftilde-EW-exact}
\end{equation}
which is used for comparison with the numerical solutions.

 We now integrate the NLO flow equations starting from a very small value of $\hat g_\Lambda$, typically
 $10^{-6}$. During the flow, $\hat g_\kappa$ grows monotonically to reach the KPZ fixed point. We focus on the beginning of the flow, before $\hat g_\kappa$ exceeds typically $10^{-2}$, which is expected to be dominated by the EW fixed point. As shown in \fref{fig:NLO-EW}, the evolution of the function  $\hat{f}_\kappa(\hat \varpi, \hat p)$ during this part of the flow is almost negligible, and one obtains a correlation function which essentially keeps its bare form. The dynamical exponent is the bare one $z=2$, as evidenced by the scaling of the half-frequency. The corresponding dynamics is purely diffusive, not affected yet by the non-linearity. We show in \fref{fig:NLO-EW} the scaling function obtained from the collapse of the data for ${C}(\hat  t,\hat  p)/C(0,\hat p)$, which exactly matches the expected EW scaling function given by \eqref{eq:ftilde-EW-exact} up to numerical precision.

\subsubsection{The IB fixed point} 
 
\begin{figure}[t]
 \tikz{
 \node at (-1,0) {\includegraphics[width=8cm]{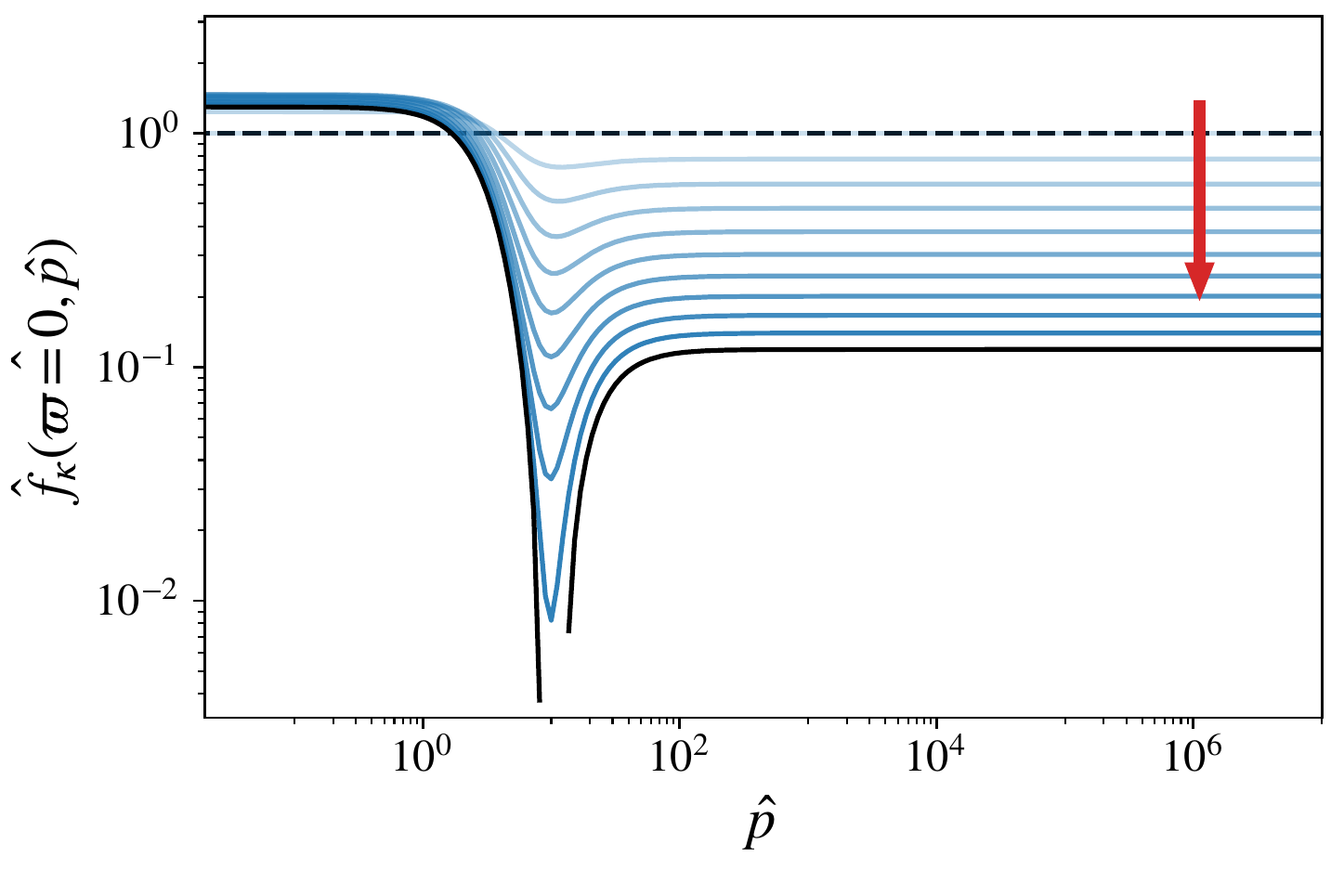}};
 \node at (8,0.5) {\includegraphics[width=8cm]{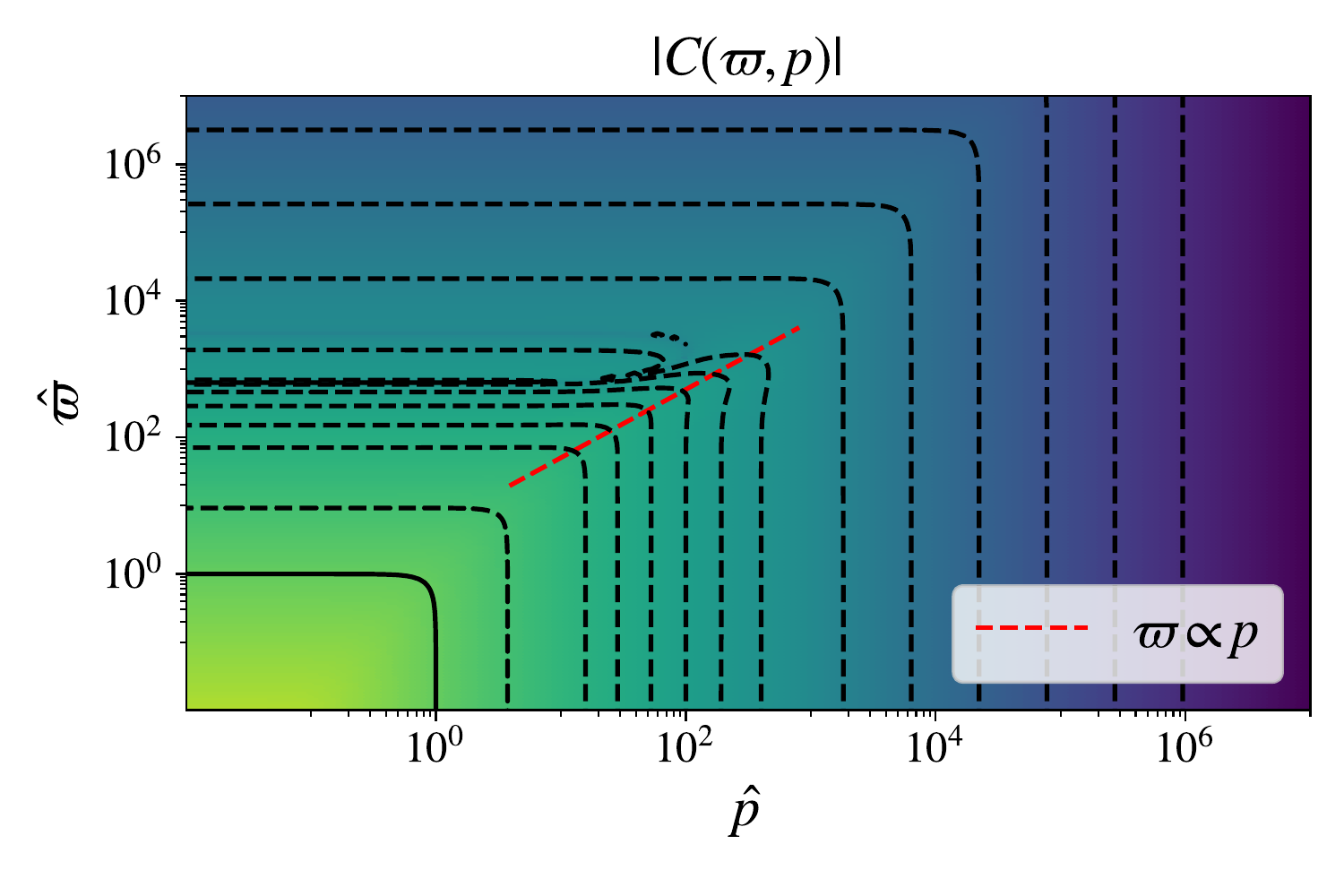}};
 \node at (-1,-5.5) {\includegraphics[width=8cm]{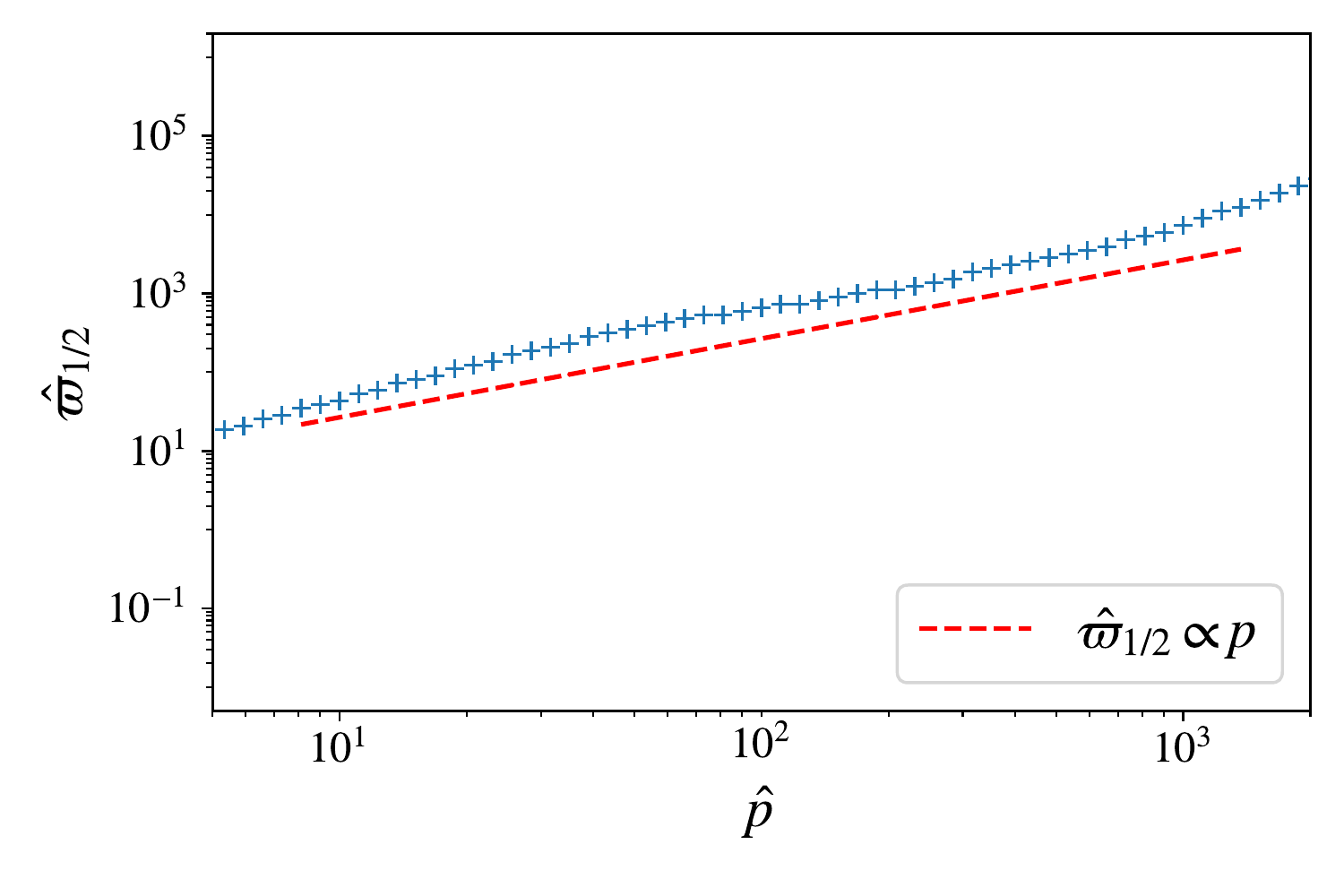}};
 \node at (8,-5) {\includegraphics[width=8cm]{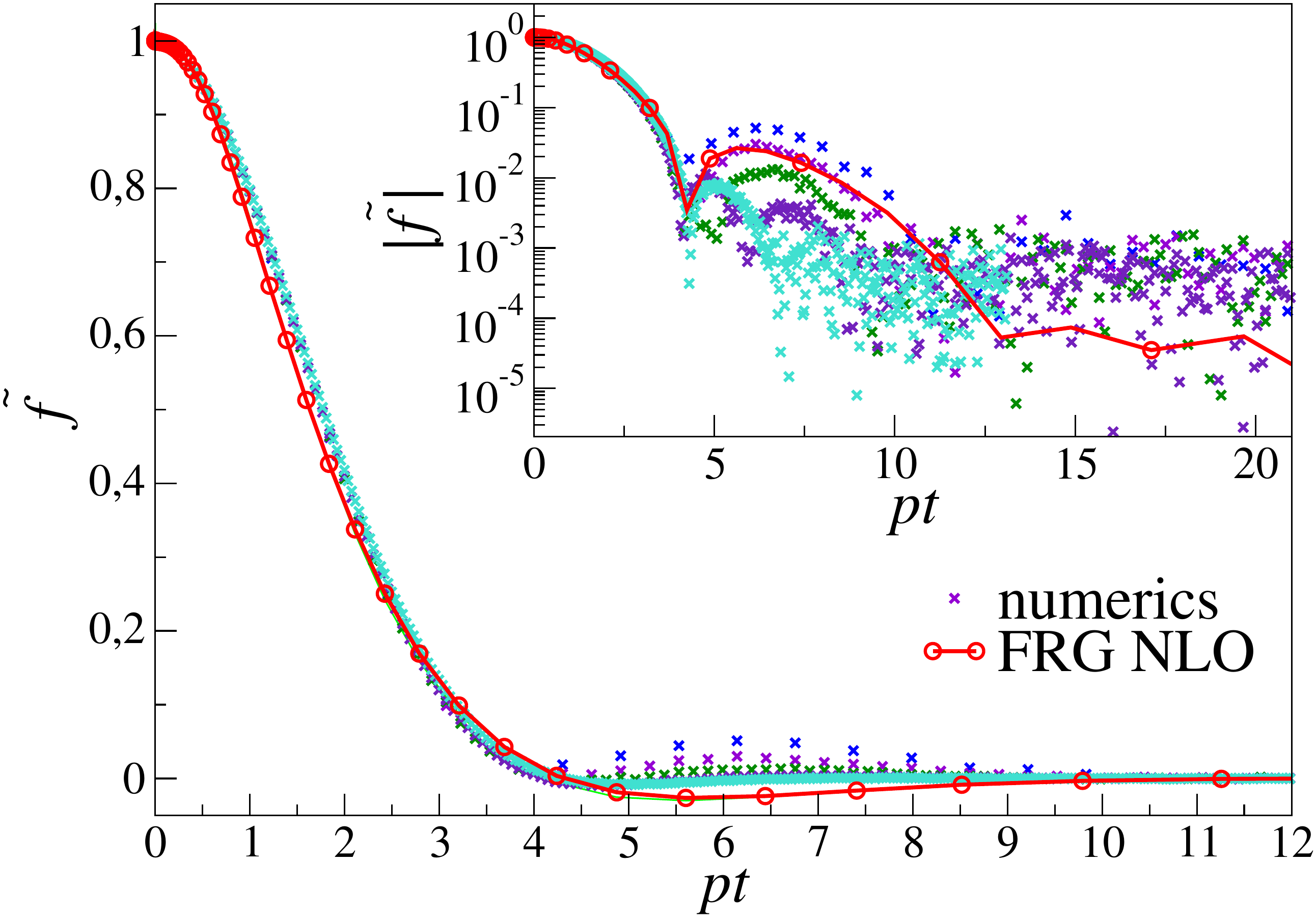}};
 } 
  \caption{({\it Top left panel}) Evolution of $\hat f_\kappa(\hat \varpi=0,\hat p)$ during the initial UV flow  from a large value of $\hat g_\Lambda =10^{3}$. ({\it Top right panel}) Correlation function $C(\hat \varpi, \hat p)$, and ({\it bottom left panel})  half-frequency  $\hat\varpi_{1/2}$ as a function of $\hat p$, which both show the IB dynamical exponent $\varpi\sim p^{z}$ with $z=1$.  ({\it Bottom right panel}) Collapse of the correlation function onto the IB scaling function, compared with  the result from the numerical simulations of Ref. \cite{Brachet2022}.}
  \label{fig:NLO-IB}
 \end{figure}

 We now integrate the NLO flow equations starting from a very large value of $\hat g_\Lambda$, typically
 $10^{3}$. During the flow, $\hat g_\kappa$ decreases monotonically. We focus on the beginning of the flow, before $\hat g_\kappa$ becomes smaller than typically $10^{2}$, where we expect it to be dominated by the IB fixed point. As shown in \fref{fig:NLO-IB}, the evolution of the function  $\hat{f}_\kappa(\hat\varpi, \hat p)$ during this part of the flow is significant, and the function develops a negative dip after a finite RG time. The 
 corresponding  correlation function shows a $z=1$ dynamical exponent, confirmed by the scaling of the half-frequency.  We show in \fref{fig:NLO-IB} the scaling function obtained from the collapse of the data for ${C}(\hat t, \hat p)/C(0,\hat p)$, which is in close agreement with the data from the numerical simulations of \cite{Brachet2022}, at least from $pt=0$ to   $pt\simeq 4$. It reproduces in particular the negative dip observed in the data from the  simulations. Both  the data from the numerical simulations and  from the FRG at NLO saturate at a numerical error level $\sim 10^{-4}-10^{-5}$ beyond this value. Hence, they are not precise enough to resolve the behavior of the function at large time, and in particular to uncover the crossover to an exponential decay predicted by the exact asymptotic FRG solution \eqref{eq:ftilde-IB-exact}.

 \section{Exact flow equations in the large wavenumber limit}
 \label{app:largep}
 
 In this section, we derive the exact FRG flow equation for the 2-point correlation function in the limit of large wavenumber, and solve it at the fixed point. The calculations are similar to \cite{Canet2016,Tarpin2018,Canet2022}, to which we refer the reader for further details. 
 This calculation  rigorously proves that $z=1$ at the IB fixed point, and provides the exact asymptotic form of the IB scaling function. 
We first derive the Ward identities associated with extended symmetries, and then show how they allow one to exactly close the FRG flow equations in the limit of large wavenumbers.
In order to simply exploit the analogy with the calculations available for the Navier-Stokes (NS) equation, we consider the  Burgers action rather than the KPZ one, which reads in one dimension:
\begin{equation}
{\cal S}_{\rm Burgers}[v] = \int_{t,x}\Big\{\tilde{v}\Big[\p_t v +v \p_x v - \nu \p^2_x v   \Big] - {\cal D}\big(\p_x\tilde{v}\big)^2\Big\}\, ,
\label{eq:actionBurgers}
\end{equation}
where ${\cal D} = D \lambda^2$, and the conservative noise is inherited from the mapping with the KPZ equation. The two actions are strictly equivalent, and the results we obtain in the following apply for both the KPZ and Burgers equations.

\subsection{Ward identities associated with the extended symmetries}

Extended symmetries correspond to general field and coordinate transformations under which the action is not strictly invariant, but whose variation is linear in the fields.
Such symmetries allow one to derive functional Ward identities, from which can be deduced an infinite set of exact identities relating  the different vertex functions (or correlation functions) of the theory \cite{Canet2015}.
 These relations can then be used to constrain any approximations, or even lead to  exact results, as here.

The Burgers action shares with the NS one a first extended symmetry which is the 
time-dependent Galilean symmetry \cite{Adzhemyan99,Antonov96,berera07,Canet2010,Canet2016}, corresponding to the following infinitesimal field transformation
\begin{equation}
 \delta v(t,x)=-\dot{\epsilon}(t)+\epsilon(t) 
\partial_x v(t,x)\,, \qquad
 \delta \tilde v(t,x)=\epsilon(t) \partial_x \tilde 
v(t,x)\, ,
\label{defG}
\end{equation}
where $\dot{\epsilon} = \p_t \epsilon$. When 
$\epsilon(t)\equiv \epsilon$ is an arbitrary constant,   the 
transformation corresponds to a translation in space, and  when 
$\epsilon(t)\equiv  \epsilon\times t$ it corresponds to the usual 
 Galilean transformation. The transformation \eqref{defG} for an arbitrary infinitesimal time-dependent function $\epsilon(t)$ realizes an extended symmetry, which leads to the functional Ward identity
 \begin{equation}
\label{wardgalilee}
 \int_{x} \Big\{\big(\p_t +  \partial_x 
u(t,x)\big)\frac{\delta \Gamma_\kappa}{\delta u(t,x)}   \nonumber
+ \partial_x \tilde u(t,x) \frac{\delta \Gamma_\kappa}{\delta \tilde
u(t,x)}\Big\}= -\int_{x} \p_t^2 \tilde u(t,x)\, ,
\end{equation}
where $u=\langle v \rangle$, and  $\tilde u=\langle \tilde v \rangle$.
 By taking functional derivatives of this identity with respect to fields $u$ or $\tilde u$, one deduces the following general identities
 \begin{equation}
\Gamma_\kappa^{(m+1,n)}\Bigr(\cdots,\underbrace{\omega_\ell,p_\ell=0}_{\ell={\rm velocity \;index}},\cdots\Bigr)  =
-\sum_{i=1}^{m+n}\frac{p_{i}}{\omega_\ell}\Gamma_\kappa^{(m,n)}\Bigr(\cdots,\underbrace{\omega_{i}+\omega_\ell, p_{i}}_{i^{{\rm th}}{\rm field}},\cdots\Bigr)\, .
\label{eq:wardGalNtheta}
\end{equation}
These identities imply that any $(m+1)$-point vertex with one zero wavevector carried by a velocity field can be expressed exactly in terms of lower-order $m$-point vertices.
In particular, we will employ the following identities giving $\bar\Gamma_\kappa^{(3)}$ (using implicit momentum and frequency conservation) with a zero wavevector carried by a velocity field 
\begin{align}
 \bar\Gamma_\kappa^{(2,1)}(\omega,& q=0,\varpi,
p)=-\frac{p}{\omega}\Big(\bar\Gamma_\kappa^{(1,1)}(\varpi+\omega,  
p)-\bar\Gamma_\kappa^{(1,1)}(\varpi,  p)\Big)\label{ward-21}\\
 \bar\Gamma_\kappa^{(1,2)}(\omega,& q= 0,\varpi,  
p)=-\frac{p}{\omega}\Big(\bar\Gamma_\kappa^{(0,2)}(\varpi+\omega,  
p)-\bar\Gamma_\kappa^{(0,2)}(\varpi,  p)\Big)\, ,\label{ward-12}
\end{align}
and the ones giving $\bar\Gamma_\kappa^{(4)}$ with two vanishing wavevectors carried by  velocity fields (obtained by applying twice \eqref{eq:wardGalNtheta})
\begin{align}
\bar\Gamma_\kappa^{(2,2)}(\omega,  q =   0, -\omega, -  
q =   0,\varpi,   p) &= \frac{p^2}{\omega^2}\Bigg[ 
\bar\Gamma_\kappa^{(0,2)}(\varpi+\omega,  p) -2 
\bar\Gamma_\kappa^{(0,2)}(\varpi,  p) 
+\bar\Gamma_\kappa^{(0,2)}(\varpi-\omega,  p)  \Bigg] \label{ward-22} \\
\bar\Gamma_\kappa^{(3,1)}(\omega,  q =   0, -\omega, -  
q =   0,\varpi,   p)& = \frac{p^2}{\omega^2}\Bigg[ 
\bar\Gamma_\kappa^{(1,1)}(\varpi+\omega,  p) -2 
\bar\Gamma_\kappa^{(1,1)}(\varpi,  p) 
+\bar\Gamma_\kappa^{(1,1)}(\varpi-\omega,  p)  \Bigg] \label{ward-31}.
\end{align}

The Burgers action admits another extended symmetry, which corresponds to the following infinitesimal time-dependent shift of the response velocity field
\begin{equation}
\delta \tilde v(t,x) = \tilde \epsilon(t)\, .
\label{eq:shift}
\end{equation}
Let us emphasize that the NS action also admits a time-dependent shift symmetry in the response field sector (in the form of a joint shift of the response velocity and response pressure fields), unveiled in \cite{Canet2015}, which is crucially rooted in incompressibility. Indeed, this extended symmetry relies on a compensation between the variation of the advection term and the  variation of the term encoding incompressibility \cite{Canet2015}. Since the Burgers equation is devoid of pressure, a shift of the response velocity field cannot be absorbed by a simultaneous shift of the response pressure, and thus the Burgers action  is not invariant in general dimension  under such a shift. However, in 1D, the advection term can be written as  $\tilde v v\p_x v = \tilde v(\p_x v^2)/2$, such that this term turns out to be invariant under the shift \eqref{eq:shift}, and the overall variation
of the Burgers action is linear in the field. The corresponding functional Ward identity simply writes
\begin{equation}
\int_{x}\frac{\delta \Gamma_\kappa}{\delta \tilde u(t,x)} = 
\int_{x}  \p_t u(t,x)\, .
\label{eq:Wardshift}
\end{equation}
By taking one functional derivative of this identity with respect to $u$, and  Fourier transforming, one obtains
\begin{equation}
 \bar \Gamma_\kappa^{(1,1)}(\omega,p= 0)=i\omega\, .
 \label{gam11ward}
\end{equation}
By further  functional differentiations with respect to velocity and response fields, one deduces the infinite set of exact identities
\begin{equation}
\bar\Gamma_\kappa^{(m,n)}\Bigr(\cdots,\underbrace{\omega_\ell,p_\ell=0}_{\ell={\rm response \;velocity \;index}},\cdots\Bigr)  = 0\,,
\label{eq:wardshiftN}
\end{equation}
which implies that any vertex of order $m+n>2$ with one zero wavevector carried by a response velocity field vanishes.

In one dimension,  the Burgers action also admits the discrete time reversal symmetry, which corresponds to the following transformation \cite{Canet2010}
 \begin{equation}
 v(t,x)= - v(-t,x) \,, \qquad
 \tilde v(t,x)=  \tilde v(-t,x) - \frac{\nu}{{\cal D}}  v(-t,x)\, .
\label{eq:defTSR}
\end{equation}
It yields the functional identity
\begin{equation}
\Gamma_{\kappa}[v(t,x),\tilde v(t,x)] = \Gamma_{\kappa}[- v(-t,x),\tilde v(-t,x) - \frac{\nu}{{\cal D}} v(-t,x)]\, . 
\end{equation} 
Taking functional derivative of this identity, Fourier transforming and using parity, one obtains for the 2-point functions 
\begin{equation}
2\Re {\rm e}\Big[\bar{\Gamma}^{(1,1)}_\kappa(\omega,q)\Big] = - \dfrac{\nu}{{\cal D}}   \bar{\Gamma}^{(0,2)}_\kappa(\omega,q)\, .
\end{equation}

\subsection{Exact closure in the limit of large wavenumbers}

It turns out that the FRG flow equations can be controlled exactly in the limit of large wavenumbers  thanks to the presence of the regulator, which was first shown within the BMW framework \cite{Blaizot2006,Blaizot2007,Benitez2012}, and thoroughly used in the context of turbulence to obtain exact results for generic $n$-point correlation functions in the limit of large wavenumbers \cite{Canet2017,Tarpin2018,Tarpin2019,Canet2022}.
As explained in the main paper,  the presence of the derivative of the regulator in the flow equation cuts the internal momentum to values $|q|\lesssim \kappa$. Thus, in the limit of large external momentum $p\gg \kappa$, one has $|q|\ll p$, and the vertices in the flow equation ({\it eg.} Eq.~\eqref{eq:dsgam2} for the 2-point functions) can be expanded in powers of $q$. Indeed, the analyticity of all the vertices at any finite $\kappa$ is ensured both in the IR and in the UV by the presence of the regulator, such that they can be safely expanded \cite{Dupuis2021}. One can show that this expansion is asymptotically exact in the limit $p\to \infty$ \cite{Benitez2012}.
The condition of large wavenumber corresponds to $p\gg \kappa$, and the RG scale $\kappa$ ultimately tends to zero, or to the inverse of a large scale  beyond which fluctuations are negligible ({\it eg.} the integral scale $L$ for turbulence). Thus the large $p$ region is expected to start at moderate values of $p$, within the universal regime, well before the UV cutoff.
 In fact, it was shown in numerical simulations of  Navier-Stokes turbulence that the large $p$ regime starts when the direct energy transfer from a forcing mode to an inertial range mode becomes negligible, which is achieved for a wavenumber smaller but not too far from the inverse integral scale \cite{Gorbunova2021}. Thus, the large $p$ regime encompasses wavenumbers within the inertial range  down to the dissipative range.   For the Burgers equation, in order to describe the UV  sector, the relevant scale is the one separating the IR and the UV momenta, that we denote $k_c$. 
 
We consider the flow equations for the 2-point functions $\bar{\Gamma}_\kappa^{(1,1)}$ and $\bar{\Gamma}_\kappa^{(0,2)}$, given by the exact equation
 \eqref{eq:dsgam2} and  represented in \fref{fig:flowG02} and \fref{fig:flowG11}. Our aim is to show that each of the diagrams involved in these equations is 
either negligible, or closed (expressed in terms of 2-point functions only) in the 
large $p$ limit.
 Indeed, in this limit, one can set $q=0$ in all the vertices, where $q$ is the internal wavevector, circulating in the loop. 
  If this wavevector enters a vertex on a $\tilde u$ leg (represented by an outgoing arrow), then the corresponding vertex vanishes because of  the Ward identity \eqref{eq:wardshiftN}, and the whole diagram does not contribute to the flow equation. This is the case for the 
 diagrams $(b)$, $(d)$ and $(e)$ both in \fref{fig:flowG02} and in \fref{fig:flowG11}, which can be neglected in the large $p$ limit. 
For the remaining diagrams, the $q=0$ wavevector enters each vertex on a $u$ leg (represented by an ingoing arrow). Such vertex can be expressed in terms of 2-point functions using the Ward identities related to the extended Galilean symmetry, as we now establish.

More precisely, to calculate a given diagram, one first write down all momentum configurations generated by the matrix product in \eqref{eq:dsgam2}, and then distribute in each the $\tilde \p_s$ operator. This yields either a $\tilde \p_s \bar C_\kappa(Q)$ or  a $\tilde \p_s \bar R_\kappa(Q)$, where $Q=\pm q$ or $Q=\pm (p+q)$. When $Q=\pm (p+q)$, one can change variables in the integrals $(q' = \mp (p+q),\omega' =  \mp (\varpi+\omega))$ such that it is always the momentum $q$ which is cutoff in the loop. Even in the remaining diagrams, some terms turn out to vanish  because of \eqref{eq:wardshiftN}, that is the $q$ momentum appears to enter a vertex on a $\tilde u$ leg. We only detail below the non-zero terms in all the remaining diagrams.

\vspace{0.2cm}
\noindent{\bf Flow of $\bar{\Gamma}_\kappa^{(1,1)}$}
\vspace{0.2cm}

The diagram $(a)$ of \fref{fig:flowG11} can be expressed, in the large $p$ limit and using  \eqref{ward-31}, as
\begin{align}
\left[\partial_s \bar \Gamma_\kappa^{(1,1)}(\varpi, 
  p)\right]_{(a)}&=\frac 1 2  
\int_{\omega,  q} \bar \Gamma_\kappa^{(3,1)}(\omega,  q,-\omega,   
-q, \varpi,  p) \tilde\partial_s  \bar C_{\kappa}(\omega,  q) \nonumber \\
&\stackrel{p\to \infty}{=}\frac 1 2 p^2 \int_\omega \frac 1 {\omega^2}\Bigg[ 
\bar \Gamma^{(1,1)}_{ \kappa}(\omega+\varpi,  p) -2 \bar \Gamma^{(1,1)}_{ \kappa}(\varpi,  p) 
+\bar \Gamma^{(1,1)}_{ \kappa}(-\omega+\varpi,  p)  \Bigg] \;\tilde \p_s\int_{  q} 
\bar C_{\kappa}(\omega,  q)\, .
\label{diag1A}
\end{align}
The non-zero contribution of diagram (c)  can be written as
\begin{align}
\left[\partial_s \bar \Gamma_\kappa^{(1,1)}(\varpi,   p)\right]_{(c)}
 =&- \int_{\omega,  q} \bar \Gamma_\kappa^{(2,1)}(\omega,  q,\varpi,  
p)\bar R_{\kappa}(-\omega-\varpi,  -p-q)
 \bar \Gamma_\kappa^{(2,1)}(\omega+\varpi,  p+q,-\omega,  -q)\;\tilde\partial_s 
\bar C_{\kappa}(\omega,  q)\nonumber\\
&\stackrel{p\to \infty}{=} - p^2 \int_{\omega} 
\left[\frac{\bar\Gamma^{(1,1)}_\kappa(\omega+\varpi,  p)- \bar\Gamma^{(1,1)}_\kappa(\varpi, 
p)}{\omega}\right]^2 \bar R_{\kappa}(-\omega-\varpi, p) 
\;\tilde\partial_s \int_{ q} \bar C_{\kappa}(\omega,q)\,,
\label{diag1C}
\end{align}
 where we replaced the vertices in the first line by their large $p$ limit given by \eqref{ward-31}. Summing up the two contributions 
 \eqref{diag1A} and \eqref{diag1C}, we obtain
\begin{align}
\partial_s \bar\Gamma^{(1,1)}_{\kappa}(\varpi,  p)&=  p^2 
\int_{\omega}   \Bigg\{ -\left[\frac{\bar \Gamma^{(1,1)}_\kappa(\omega+\varpi, p)- 
\bar \Gamma^{(1,1)}_\kappa(\varpi,  p)}{\omega}\right]^2  \bar R_\kappa(-\omega-\varpi, p) \nonumber\\
& + \frac 1 {2 \omega^2}\Bigg[\bar \Gamma^{(1,1)}_{\kappa}(\varpi+\omega, p) -2 
\bar\Gamma^{(1,1)}_{\kappa}(\varpi,  p) +\bar\Gamma^{(1,1)}_{\kappa}(\varpi-\omega, p)  
\Bigg] \Bigg\}
\times \tilde\partial_s \int_{q} \bar C_{\kappa}(\omega,q) \, .
\label{dtGam11}
\end{align}

\vspace{0.2cm}
\noindent{\bf Flow of $\bar{\Gamma}_\kappa^{(0,2)}$}
\vspace{0.2cm}

Similarly, the diagram $(a)$ of \fref{fig:flowG02} can be expressed, in the large $p$ limit and using  \eqref{ward-22}, as
\begin{align}
\left[\partial_s \bar \Gamma_\kappa^{(0,2)}(\varpi,   p) \right]_{(a)}
&=\frac 1 2\int_{\omega} 
\bar \Gamma_\kappa^{(2,2)}(\omega, q,- \omega,  -q,\varpi, p)\; 
 \int_{\omega, q} \tilde\partial_s  \bar C_{\kappa}(\omega,  q)\nonumber\\
&\stackrel{p\to \infty}{=} \frac 1 2 p^2 \int_\omega \frac 1 
{\omega^2}\left[ \bar \Gamma^{(0,2)}_ \kappa(\omega+\varpi,  p) -2 
\bar \Gamma^{(0,2)}_ \kappa(\varpi,  p) +\bar \Gamma^{(0,2)}_ \kappa(-\omega+\varpi,  p)  
\right]\; \tilde\partial_s \int_{  q} \bar C_{ \kappa}(\omega,  q)\, . 
\label{diag2A}
\end{align}
The expression for the diagram $(c)$ is obtained using \eqref{ward-21} for the two vertices, and reads
\begin{align}
 \left[\partial_s \bar \Gamma_\kappa^{(0,2)}(\varpi,   p)\right]_{(c)} & =- 
\int_{\omega,q} \bar \Gamma_\kappa^{(2,1)}(\omega,  q,-\omega-\varpi,-  p-q)\bar C_{\kappa}(\omega+\varpi,  p+q)\bar \Gamma_\kappa^{(2,1)}(\omega+\varpi,  
p+q,-\omega,  -q)
\; \tilde\partial_s \bar C_{\kappa}(\omega,  q)\nonumber\\
 &\stackrel{p\to \infty}{=}  - p^2 \int_{\omega} \frac 1 
{\omega^2}\left(\bar \Gamma^{(1,1)}_\kappa(-\varpi,   p)- 
\bar \Gamma^{(1,1)}_ \kappa(-\varpi-\omega,   p)\right)\nonumber \\ 
 &\times \left(\bar \Gamma^{(1,1)}_\kappa(\varpi,   p)- 
\bar \Gamma^{(1,1)}_ \kappa(\varpi+\omega,   p)\right) 
\bar C_{ \kappa}(\omega+\varpi,  p) \; \tilde\partial_s \int_{  q} 
\bar C_{ \kappa}(\omega,  q)\,.
\label{diag2C}
\end{align}
Finally, the expression for 
 the diagram $(f)$ is obtained using \eqref{ward-21} and \eqref{ward-12} for the  vertices, and reads
\begin{align}
\left[\partial_s \bar \Gamma_\kappa^{(0,2)}(\varpi,   p)\right]_{(f)}
 &= -  \int_{\omega,  q} \bar \Gamma_\kappa^{(1,2)}(\omega,  q,\varpi,  p)\bar R_\kappa(-\omega-\varpi,  -p-q)
 \bar \Gamma_\kappa^{(2,1)}(\omega+\varpi,  p+q,-\omega,  -q)\;\tilde\partial_s 
\bar C_\kappa(\omega,  q)+c.c.\nonumber\\
&\stackrel{p\to \infty}{=}  -  p^2 \int_{\omega}
\left[\frac{\bar \Gamma^{(0,2)}_\kappa(\omega+\varpi,   p)- \bar \Gamma^{(0,2)}_\kappa(\varpi,  p)}{\omega}\right] \times
\left[\frac{\bar \Gamma^{(1,1)}_\kappa(\omega+\varpi,   p)- \bar \Gamma^{(1,1)}_\kappa(\varpi, 
  p)}{\omega}\right] \nonumber\\
 &\times \bar R_{ \kappa}(-\omega-\varpi,  p) \;\tilde\partial_s \int_{  
q} \bar C_{ \kappa}(\omega,  q)+c.c.\, ,\label{diag2F}
\end{align}
where $c.c.$ denotes the complex conjugate.
Adding the three contributions \eqref{diag2A}, \eqref{diag2C} and \eqref{diag2F} finally leads to
\begin{align}
\partial_s \bar\Gamma^{(0,2)}_{\kappa}(\varpi,  p)&=  p^2 
\int_{\omega}   \Bigg\{-
 \Bigg|\frac{\bar\Gamma^{(1,1)}_\kappa(\varpi,  p)- \bar\Gamma^{(1,1)}_\kappa(\varpi+\omega,  p)}{\omega} \Bigg|^2 \,    \bar C_{\kappa}(\omega+\varpi, p) \nonumber\\
&- 2 \left[\frac{\bar \Gamma^{(0,2)}_ \kappa(\omega+\varpi,   p)- 
\bar \Gamma^{(0,2)}_ \kappa(\varpi,   p)}
{\omega}\right] \times \Re \left\{\left[\frac{\bar \Gamma^{(1,1)}_ \kappa(\omega+\varpi, 
  p)- \bar \Gamma^{(1,1)}_ \kappa(\varpi,   p)}{\omega}\right]  \bar R_{ \kappa}(-\omega-\varpi,  p) \right\}\nonumber \\
&+ \frac 1 {2\omega^2}\Bigg[ \bar \Gamma^{(0,2)}_ \kappa(\omega+\varpi,  p) -2 
\bar \Gamma^{(0,2)}_ \kappa(\varpi,  p) +\bar \Gamma^{(0,2)}_ \kappa(-\omega+\varpi,  p)  
\Bigg] \Bigg\}  \times \tilde\partial_s \int_{  q} \bar C_{ \kappa}(\omega,  q). 
\label{dtGam02}
\end{align}

The two flow  equations \eqref{dtGam11} and \eqref{dtGam02} for the 2-point functions are thus closed in the limit of large $p$, but they are nonlinear. It turns out that they endow a much simpler form  when expressed for the correlation functions rather than for the $\bar \Gamma_\kappa^{(2)}$.
Indeed, using the definition \eqref{eq:defG}, one can calculate the flow equations for $\bar C_\kappa$ and $\bar R_\kappa$, and one finds, using parity and after some algebra 
\begin{align}
\p_s \bar C_\kappa(\varpi, p) &=  p^2 
\int_{\omega} \frac{1}{2\omega^2}\Big[\bar C_\kappa(\omega+\varpi,  p) -2 
\bar C_ \kappa(\varpi,  p) +\bar C_\kappa(-\omega+\varpi,  p) \Big] \tilde\partial_s \int_{  q} \bar C_{ \kappa}(\omega,  q)\nonumber\\
\p_s \bar R_\kappa(\varpi, p) &=  p^2 
\int_{\omega} \frac{1}{2\omega^2}\Big[\bar R_\kappa(\omega+\varpi,  p) -2 
\bar R_\kappa(\varpi,  p) +\bar R_\kappa(-\omega+\varpi,  p) \Big] \tilde\partial_s \int_{  q} \bar C_{ \kappa}(\omega,  q)\, .
\end{align}
Note that under this form, it is straightforward to check that these two equations preserve along the flow the time-reversal symmetry which reads for the correlation and response functions 
\begin{equation}
2\Re {\rm e}\Big[\bar R_\kappa(\varpi,q)\Big] = \dfrac{\nu}{{\cal D}}   \bar C_\kappa(\varpi,q)\, .
\end{equation}
We now focus only on the equation for $\bar C_\kappa$. 
Fourier transforming back to real time, one obtains
\begin{equation}
\p_s \bar C_\kappa(t, p) =  -p^2 \bar C_\kappa(t,p) \int_\omega \dfrac{\cos(\omega t)-1 }{\omega^2} J_\kappa(\omega)\, , \quad J_\kappa(\omega) =-\tilde\partial_s \int_{  q} \bar C_{ \kappa}(\omega,  q)\, .
\label{eq:dsC}
\end{equation}

\subsection{Fixed-point solution and IB scaling function}

We can now derive the solution of \eqref{eq:dsC} at the IB fixed point.  
We introduce as before dimensionless variables $\hat p = p/\kappa$, $\hat t = t \kappa^{2-\eta_\kappa}$, and  define the dimensionless correlation function via
$\bar C_\kappa( t, p) = \kappa^{-4+\eta_\kappa}\hat C_\kappa(\hat t = t \kappa^{2-\eta_\kappa},\hat p = p/\kappa)$.
Its flow equation is given by
\begin{equation}
\Big[\p_s -(4-\eta_\kappa) -\hat p\p_{\hat p} + (2-\eta_\kappa){\hat t \p_{\hat t}} \Big]\hat C_\kappa(\hat t, \hat p) =  -\hat p^2 \hat C_\kappa(\hat t, \hat p) \int_\omega \dfrac{\cos(\hat \omega \hat t)-1 }{\hat \omega^2} \hat {J}_\kappa(\hat \omega)\,  .
\end{equation}
The fixed-point equation corresponds to 
\begin{equation}
\Big[ -(4-\eta_*) -\hat p\p_{\hat p} + (2-\eta_*){\hat t \p_{\hat t}} \Big]\hat C_*(\hat t, \hat p) = - \hat p^2 \hat C_*(\hat t, \hat p) \int_\omega \dfrac{\cos(\hat \omega \hat t)-1 }{\hat \omega^2} \hat {J}_*(\hat \omega)\,  .
\label{eq:dsChat}
\end{equation}
 This equation can be simplified upon defining $\hat{C}_*(\hat{t},\hat{p}) = \hat{p}^{4-\eta_*} \hat H(\hat p,\hat{y})$ with $\hat{y}=\hat t \hat{p}^{z_*}$. The fixed-point equation for $\hat H$ writes
\begin{equation}
\p_{\hat p} \hat H(\hat p,\hat{y}) = \hat p \hat H(\hat p, \hat y) \int_\omega \dfrac{\cos(\hat y\hat \omega/p^{z_*} )-1 }{\hat \omega^2} \hat {J}_*(\hat \omega)\,.
\label{eq:dpH}
\end{equation}
The explicit solution can be obtained in the two limits of small and large time delays, 
which allow one to simplify the integrals \cite{Tarpin2018}.
Indeed, at small time delays, one can expand the cosine in the integral, yielding
\begin{equation}
\p_{\hat p} \hat H(\hat p,\hat{y}) = -\hat p^{1-2 z_*} \hat y^2 \hat \alpha_0 \hat H_*(\hat p, \hat y) \, , \quad \hat \alpha_0 = \frac 1 2\int_\omega \hat {J}_*(\hat \omega)\,.
\end{equation}
The solution is given by
\begin{equation}
\hat H(\hat p,\hat{y}) = F_0(\hat{y})\,\exp\big(-\hat \mu_0 \hat p^2 \hat t^2\big)\, ,
\end{equation}
with $\hat \mu_0 =\hat\alpha_0/(2-2z_*)$.
At large time delays, one can rewrite $\hat J_*(\hat\omega)$ in the integral as $\hat J_*(\hat\omega) = (\hat J_*(\hat\omega )-\hat J_*(0))+\hat J_*(0)$.
Since $\hat J_*$ is a regular even function of $\hat\omega$,
  the term $F(\hat \omega)=(\hat J_*(\hat\omega )-\hat J_*(0))/\hat\omega^2$ is an analytic function of $\hat\omega$.
 It follows that its Fourier transform $\int_{\hat\omega} \cos(\hat\omega \hat x) F(\hat\omega)$  decays exponentially in $\hat x$,
 and its integral $\int_{\hat\omega} F(\hat\omega)$ is a constant independent of $\hat x$.
 At large $t$, the integral in \eqref{eq:dpH} is thus dominated by the remaining term 
  which writes
\begin{equation}
\hat J_*(0) \int_{-\infty}^\infty\frac{d\hat \omega}{2\pi}\, \frac{\cos(\hat y\hat \omega/p^{z_*} )-1}{\hat\omega^2} = -\frac{\hat J_*(0)}{2} \left|\dfrac{\hat y}{\hat p^{z_*}}\right|\,.
\end{equation}
The fixed point equation is then
\begin{equation}
\p_{\hat p} \hat H(\hat p,\hat{y}) = -\hat p^{1- z_*} |\hat y| \hat \alpha_\infty \hat H_*(\hat p, \hat y) \, , \quad \hat \alpha_\infty = \frac{\hat J_*(0)}{2}\,.
\end{equation}
The solution is given by
\begin{equation}
H(\hat p,\hat{y}) = F_\infty(\hat{y})\,\exp\big(-\hat \mu_\infty \hat p^2 | t| \big)\,,
\end{equation}
with $\hat \mu_\infty =\hat\alpha_\infty/(2-z_*)$. Neglecting the dependence in $\hat y$ which is expected to be sub-dominant compared to $\hat p^2$ leads to the solution for the dimensionfull correlation function
 \begin{equation}
C(t,p) = C(0,p)\times \left\{  \begin{array}{l l}
 \exp\Big(-\mu_0 \,\big( pt\big)^2 \Big) & t\ll\tau_c\\ 
  \exp\Big(-\mu_\infty \,p^2 | t|\Big)  & t\gg\tau_c
 \end{array}\right.,
 \label{eq:ftilde-IB-exact}
\end{equation}
with $\mu_0 = \hat \mu_0/(k_c\tau_c)^2$ and  $\mu_\infty = \hat \mu_\infty/(k_c^2\tau_c)$ where $k_c$ is the typical momentum scale where the UV flow crosses over to the IR one, and $\tau_c=(\nu k_c^{2-\eta_*})^{-1}$ the associated crossover time scale. Both expressions are exact in the large $p$ limit. The small time  expression thus  proves the $z=1$  scaling in the UV and provides the asymptotic form of the associated scaling function, which is a Gaussian in the variable $pt$. The large time expression predicts a crossover to a new regime, which could be probed in numerical simulations with a higher resolution.

\end{widetext}

%

\end{document}